\documentclass{article}

\usepackage{hyperref}
\usepackage[a4paper, top=25mm, left=10mm, right=10mm, bottom=30mm, headsep=10mm, footskip=12mm]{geometry}
\usepackage{authblk}

\usepackage[nointlimits]{amsmath}
\usepackage{siunitx}

\usepackage{caption}
\usepackage{graphicx}
\usepackage{subfigure}
\usepackage{xcolor}

\usepackage{natbib}

\definecolor{mycolor_blue}{RGB}{66,124,161}
\definecolor{mycolor_grey}{RGB}{198,198,198}

\usepackage{tikz}
\usepackage{pgfplots}

\usetikzlibrary{patterns}
\pgfplotsset{
	kurze Legende/.style={
		legend image code/.code={
			\draw[##1,mark repeat=2,line width=0.6pt]
			plot coordinates {
				(0cm,0cm)
				(0.3cm,0cm)
			};
		}
	}
}

\pgfplotsset{
	compat = newest,
	scale only axis, 
	max space between ticks = 50pt,
	ticklabel style = {font=\footnotesize},
	legend style =  {font=\footnotesize},
	grid = major,
	grid style = {dotted},
	legend columns=1, 
	xtick pos=left,
	ytick pos=left
}

\newcommand{\longPicture}{
	\pgfplotsset{  
		width=0.8\textwidth,
		height=0.2\textwidth,
		ylabel style={text width=0.2\textwidth,align=center}
	}
}

\newcommand{\longBigPicture}{
	\pgfplotsset{  
		width=0.8\textwidth,
		height=0.3\textwidth,
		ylabel style={text width=0.2\textwidth,align=center}
	}
}

\newcommand{\smallPicture}{
	\pgfplotsset{  
		width=0.35\textwidth,
		height=0.35\textwidth,
		ylabel style={text width=0.2\textwidth,align=center}
	}
}

\newcommand{\analytiSolutionPictures}{
	\pgfplotsset{  
		width=0.24\textwidth,
		height=0.25\textwidth,
		ylabel style={text width=0.2\textwidth,align=center}
	}
}

\pgfplotsset{select coords between index/.style 2 args={
		x filter/.code={
			\ifnum\coordindex<#1\fi
			\ifnum\coordindex>#2\fi
		}
}}

\definecolor{color1}{HTML}{0060AD} % blau
\definecolor{color2}{HTML}{FF4500} % rot
\definecolor{color3}{HTML}{FFA500} % gelb
\definecolor{color4}{HTML}{006400} % gruen
\definecolor{color5}{HTML}{9400D3} % lila
\definecolor{color6}{HTML}{800000} % bordeauxrot/braun
\definecolor{color7}{HTML}{000000} % schwarz
\definecolor{color8}{HTML}{0000FF} % blau heller
\definecolor{color9}{HTML}{FF0000} % rot heller
\definecolor{mycolor_blue}{RGB}{66,124,161}% niklas blue
\definecolor{mycolor_grey}{RGB}{198,198,198} % niklas grey

\tikzstyle{line1} = [color=color7,semithick] 
\tikzstyle{line2} = [color=color2,densely dotted,semithick]
\tikzstyle{line3} = [color=color1,densely dashed,semithick]
\tikzstyle{line4} = [color=color5,dash dot,semithick]
\tikzstyle{line5} = [color=color4,dash dot dot,semithick]
\tikzstyle{line6} = [color=color6,semithick]

\tikzstyle{mark1} = [color=color7,mark=x,mark size=2pt,mark options=solid,semithick] 
\tikzstyle{mark2} = [color=color2,mark=o,mark size=2pt,mark options=solid,semithick]
\tikzstyle{mark3} = [color=color1,mark=*,mark size=2pt,mark options=solid,semithick]
\tikzstyle{mark4} = [color=color5,mark=triangle,mark size=2pt,mark options=solid,semithick]
\tikzstyle{mark5} = [color=color4,mark=square,mark size=2pt,mark options=solid,semithick]

\begin{document}

\title{Cahn-Hilliard Navier-Stokes Simulations for Marine Free-Surface Flows}

\author[1]{Niklas K\"uhl\thanks{niklas.kuehl@tuhh.de}}
\author[2]{Michael Hinze}
\author[1]{Thomas Rung}

\affil[1]{Hamburg University of Technology, Institute for Fluid Dynamics and Ship Theory, Am Schwarzenberg-Campus 4, D-21075 Hamburg, Germany}
\affil[2]{Universit\"at Koblenz-Landau, Campus Koblenz, Department of Mathematics, Universit\"atsstrasse 1, D-56070 Koblenz}

\maketitle

\begin{abstract}

The paper is devoted to the simulation of maritime two-phase flows of air and water. Emphasis is put on an extension of the classical Volume-of-Fluid (VoF) method by a diffusive contribution derived from  a Cahn-Hilliard (CH) model and its benefits for simulating immiscible, incompressible two-phase flows.
Such flows are predominantly simulated with implicit VoF schemes, which mostly employ heuristic downwind-biased approximations for the concentration transport to mimic a sharp interface. This strategy introduces a severe time step restriction  and requires pseudo time-stepping of steady flows.
Our overall goal is a sound description of the free-surface region that alleviates artificial time-step restrictions, supports an efficient and robust upwind-based approximation framework and inherently includes surface tension effects when needed. 
%The approach works on continuous level, builds a bridge to specially taylored discrete (two-phase compressive) operations and makes the latter obsolet while using standardized and physically consistent numerical convection schemes.
%
The Cahn-Hilliard Navier-Stokes (CH-NS) system is verified for an analytical Couette-flow example and the bubble formation under the influence of surface tension forces. 2D Validation examples are concerned with laminar standing waves reaching from gravity to capillary scale as well as a submerged hydrofoil flow.
The final application refers to the 3D flow around an experimentally investigated container vessel at fixed floatation for $\mathrm{Re}=1.4 \cdot10^7$ and $\mathrm{Fn =0.26}$.
Results are compared with data obtained from VoF approaches, supplemented by analytical solutions and measurements. The study indicates the superior efficiency, resharpening capability and wider predictive realm of the CH-based extension for free surface flows with a confined spatial range of interface Courant numbers.

\end{abstract}


\newlength\figureheight
\newlength\figurewidth

\section{Introduction}
Many two-phase flows are characterized by immiscible fluids that feature negligible compressibility. A prominent example refers to maritime free-surface flows. Technical applications of such flows are often subjected to large interface deformations, e.g.  breaking waves. The accurate simulation of these flows requires a computational model that conserves the mass of each phase whilst preserving a sharp interface. These requirements still pose a challenge in mesh-based computational fluid dynamics.

Engineering two-phase flow simulations mostly refer to either of two {\it interface-capturing} methods \cite{ferziger2012computational}: namely the Level-Set  \cite{osher1988fronts} and the {\it Volume-of-Fluid} (VoF) approach \cite{hirt1981volume}, which both reconstruct the free surface from an indicator function. 
The Level-Set method introduced by Osher et al. \cite{osher1988fronts} utilizes a signed distance function to characterize the interface by the zero-value iso surface. The continuous distribution of the signed distance simplifies a higher-order discretization of the related transport equation, and the geometry of the interface
can be determined with improved accuracy. A drawback of the Level-Set method is that it does not guarantee mass conservation. 
Two-phase applications of the VoF method suggested by Noh et al. \cite{noh1976slic} and later refined by Hirt and Nichols  \cite{hirt1981volume} usually employ a scalar volume concentration, of a foreground phase to identify the fluid state of each cell. The method is conservative and capable to predict merging and rupturing of free surfaces. 
For immiscible fluids, any mixing of both phases is undesired but numerically difficult to avoid. 
Different strategies are conceivable to improve interface compression: Geometric reconstruction schemes, e.g. SLIC \cite{noh1976slic}, PLIC \cite{hirt1981volume} or LVIRA\cite{pilliod2004second}, and dedicated downwind-biased advection schemes, e.g. CICSAM \cite{ubbink1999method}, HRIC \cite{muzaferija1998computation}, IGDS \cite{jasak1999high} or BRICS \cite{wackers2011free}. Geometric reconstruction schemes are afflicted with a considerable algorithmic complexity which reduces their popularity.
Dedicated advection schemes are slightly heuristic but fairly simple to implement. They maintain an approximately sharp interface subject to sufficiently small time steps. On the downside, they require transient implicit simulations even for steady state problems, e.g. the calm-water resistance of steady cruising  ships. To further improve the interface compression, some authors have proposed to add an artificial compression or anti-diffusion term, e.g. \cite{so2011anti,heyns2013development}. These methods rely on heuristic compression factors and   improve the compressiveness at the expense of a reduced numerical stability.

If surface tension effects are negligible, VoF models using dedicated advection schemes are deemed a good compromise between efficiency, accuracy and conservation properties. 
An alternative, much less common approach refers to  diffuse interface models, often labelled Cahn-Hilliard (CH) models \cite{cahn1958free}. Here, the (ideally sharp) interface is replaced by a (thin) layer where the fluids mix. The approach is able to mimic phase separation and thus promises resharpening features which are attractive for  engineering simulations. Although, the neglect of surface tension is an acceptable assumption in many engineering problems, it appears that the CH approach incorporates surface tension in a natural way and no additional model, e.g. the Continuum Method \cite{brackbill1992continuum, lafaurie94}, is required.

There exist a variety of different CH approaches for two fluids, e.g. models governed by fluids with matched densities (labelled as Model H \cite{hohenberg1977theory}), identical viscosities (Boussinesq Fluid  \cite{jacqmin1999calculation}) or so-called thermodynamically consistent systems  \cite{lowengrub1998quasi,abels2012thermodynamically},  just to name a few. 
In addition to the different CH  variants, different strategies for their coupling with the momentum and continuity equations of 
have been suggested. Further distinctions refer to  balancing either the mass or the volume fluxes between both phases \cite{lowengrub1998quasi,ding07},  the considered baseline conservation equations \cite{abels2012thermodynamically,ding07} and the introduction of modifications to ensure thermodynamic consistency \cite{lowengrub1998quasi,abels2012thermodynamically}.  
The VoF scheme offers a closed system of PDEs, but entails evolved parametrized approximations. On the contrary, three additional physical parameters occur in the CH method.
The first and second correspond to the transition length as well as the surface tension coefficient, the third parameter refers to the mobility that governs the strength of the phase separation process. Combining the CH model  with the Navier-Stokes equations, essentially results in an augmented VoF formulation. This inheres a non-linear, diffusive right-hand side of order four,  
which is  zero outside the inter facial region. The non-linear character is beneficial. It supports an accurate computation of surface tension effects when the interface is adequately resolved and the use of stable, upwind-biased advective approximations in under-resolved flow simulations.
To this end, a compressive numerical method is suggested for simulations, where the  transition length is under-resolved by the numerical grid and surface tension influences cannot be displayed.
The latter is based on an automatic adjustment of the mobility parameter.  Possible, minimal blurs are bypassed  with a non-linear state equation. The resulting system is virtually insensitive for spatial and temporal resolutions aspects.

The remainder of the paper is organized as follows: Sections \ref{sec:state_equat} outlines the mathematical model including a brief introduction into the  diffuse interface model. The subsequent third section describes the numerical procedure and outlines implementation aspects. Section \ref{sec:one_dimen} covers the verification. The determination of the  mobility parameter in under-resolved flows is outlined in Section \ref{sec:appro_mobil_estim}. Sections \ref{sec:gcwave} and \ref{sec:foil} validate the CH-approach and the mobility parameter estimation against frequently studied two-dimensional test cases, i.e. standing waves in the capillary and gravity scale as well as a submerged hydrofoil flow. The comparison of results and computational efforts obtained from a CH- and a VoF-approach for a widely used 3D container ship benchmark case is depicted in Section \ref{sec:three_dimen}. Final conclusions are drawn in Section \ref{sec:outlo}. Within the publication, Einstein's summation convention is used for lower-case Latin subscripts. Vectors and tensors are defined with reference to Cartesian coordinates and dimensionless variables are consistently marked with an asterisk.

%%%%%%%%%%%%%%%%%%%%%%%%%%%%%%%%%%%%%%%%%%%%%%%%%%%
\section{Mathematical Model}
%%%%%%%%%%%%%%%%%%%%%%%%%%%%%%%%%%%%%%%%%%%%%%%%%%%
\label{sec:state_equat}
Following the work of Ding et al. \cite{ding07}, one can distinguish between mass conservative and volume conservative CH strategies. To illustrate this, we define the specie densities ($\rho^a, \rho^b$) by a simple linear equation of state, which connects them to their constant bulk densities ($\rho_a, \rho_b$), i.e. $\rho^a =c\rho_a$ and $ \rho^b=(1-c) \rho_b$. The expression  $c=V_a/V$ represents the volume concentration of the foreground phase, the respective concentration of the background phase reads $V_b/V= (V-V_a)/V=(1-c)$. 
The mass conservation of the species $a$ and $b$ are governed by 
\begin{align}
\frac{\partial \rho^{a} }{\partial t} + \frac{\partial \, v_\mathrm{i}  \rho^{a} }{\partial x_\mathrm{i}}  = \sigma^{a} \, , \qquad
\frac{\partial \rho^{b} }{\partial t} + \frac{\partial \, v_\mathrm{i}  \rho^{b} }{\partial x_\mathrm{i}}  = \sigma^{b} \, , 
\label{specie1}
\end{align}
where the $ \sigma^{a}$, $ \sigma^{b}$ denote the mass transfer rate into the species $a$ and $b$, $v_i$ refers to the velocity and 
 $x_i$ denotes the spatial Cartesian coordinates. 
Using $\rho = \rho^a + \rho^b$, an analogue continuity relation is obtained for the mixture 
\begin{align}
\frac{\partial \rho }{\partial t} + \frac{\partial \, v_\mathrm{i}  \rho }{\partial x_\mathrm{i}}  &= \sigma^{a}+\sigma^{b} \, .
\label{specie2}
\end{align}
Substituting $\rho^a$ and $\rho^b$ by $c\rho_a$ and $(1-c)\rho_b$, one can reformulate (\ref{specie1}) and obtain
\begin{align}
 \frac{\partial c }{\partial t} +  \frac{\partial \, v_\mathrm{i}  c }{\partial x_\mathrm{i}}  = \frac{\sigma^{a}}{\rho_a} \, , \qquad
-\frac{\partial c }{\partial t} -  \, \left[ \frac{\partial \, v_\mathrm{i}  c }{\partial x_\mathrm{i}}  -\frac{\partial \, v_\mathrm{i} }{\partial x_\mathrm{i}} \right] 
= \frac{\sigma^{b}}{\rho_b} \, . 
\label{specie3}
\end{align}
Summing up expressions (\ref{specie3}) yields an alternative continuity equation, which describes the volume change
\begin{align}
 \frac{\partial \, v_\mathrm{i}   }{\partial x_\mathrm{i}} 
= \frac{\sigma^{a}}{\rho^a} +\frac{\sigma^{b}}{\rho^b} \quad=  \frac{\partial}{\partial x_k} \left(  j_k^a + j_k^b \right) \, ,
\label{specie4}
\end{align}
where the volume diffusion fluxes refer to $j_k^a$ etc.. Two options are conceivable: balanced mass fluxes ($\sigma^{a} =-\sigma^{b}$) or balanced diffusion fluxes
($j_k^a = - j_k^b$). Since the interface is generally thin - particularly in the sharp interface limit, where both options are identical - the preference is related to the employed numerical method.  

Most authors opt for a volume conservative approach, i.e. $j_k^a= -j_k^b$, and employ a volume averaged velocity field. In this case, mass is only globally conserved, provided that the inter facial regions don't intersect with the domain boundaries \cite{ding07}, which might be difficult for travelling waves. The continuity expression (\ref{specie4}) conveniently simplifies to a zero velocity divergence, and the conservative momentum equations are augmented by a total mass flux term $v_i (\sigma^a +\sigma^b)$ on the r.h.s., cf.   (\ref{equ:prima_momen_with_dimenX}). 
On the contrary, assuming a mass conservative approach, i.e. $\sigma^a=-\sigma^b$, one has to account for divergence effects of the observed mass-averaged velocity.

\subsection{Momentum and Continuity}
The present research opts for a mass-averaged velocity field  $v_i$ governed by the Navier-Stokes (NS) equations supplemented by a surface tension force $f_i^{\rm ST} $ 
\begin{align}
\rho 
\frac{\partial \,  v_\mathrm{i}}{\partial t} +\rho v_\mathrm{j} \frac{\partial \,    v_\mathrm{i}}{\partial x_\mathrm{j}} & = \frac{\partial }{\partial x_\mathrm{j}} \left[  2 \mu _e S_\mathrm{ij} - \left( p + \frac{2}{3} \mu_e \frac{\partial v_\mathrm{k}}{\partial x_\mathrm{k}} \right) \delta_\mathrm{ij} \right] + \rho g_\mathrm{i} +
% \psi \frac{\partial c}{\partial x_\mathrm{i}}  
f_i^{\rm ST} \label{equ:prima_momen_with_dimenX-1} \, ,
\end{align}
Assuming $\sigma^a=-\sigma^b$, the continuity equation follows from (\ref{specie2}), and yields a conservative formulation for mass and momentum 
\begin{align}
\frac{\partial \, \rho v_\mathrm{i}}{\partial t} +\frac{\partial \,  \rho v_\mathrm{j}  v_\mathrm{i}}{\partial x_\mathrm{j}} & = \frac{\partial }{\partial x_\mathrm{j}} \left[  2 \mu_e S_\mathrm{ij} - \left( p + \frac{2}{3} \mu_e \frac{\partial v_\mathrm{k}}{\partial x_\mathrm{k}} \right) \delta_\mathrm{ij} \right] + \rho g_\mathrm{i} + 
%\psi \frac{\partial c}{\partial x_\mathrm{i}} 
f_i^{\rm ST} 
 \label{equ:prima_momen_with_dimenX} \, , \\
\frac{\partial \rho }{\partial t} + \frac{\partial \, v_\mathrm{i} \rho }{\partial x_\mathrm{i}}  &= 0 \label{equ:prima_mass_with_dimenX} \, , 
\end{align}
where  $\rho$, $\mu_e$ and $p$ refer to the density, dynamic viscosity and pressure of the mixture. The unit coordinates and the strain-rate tensor are marked by $\delta_\mathrm{ij}$ and $S_\mathrm{ij} = \frac{1}{2} ( \frac{\partial v_\mathrm{i}}{\partial x_j} + \frac{\partial v_\mathrm{j}}{\partial x_\mathrm{i}} )$. The four right-hand side terms of the momentum balance (\ref{equ:prima_momen_with_dimenX}) denote viscous, pressure, body and  surface tension forces. The pressure is a numerical property which often inheres further trace terms related to the adopted phase field model and surface tension force expression.
The framework supports laminar and Reynolds-averaged (modelled) turbulent flows (RANS). In the latter case, $v_i$ and $p$ correspond to Reynolds-averaged properties and $p$ is additionally augmented by a turbulent kinetic energy ($k$) term, i.e. $2 \rho k/3$. Along with the Bousinesq hypothesis, the dynamic viscosity $\mu_e = \mu+ \mu_t$ of turbulent flows consists of a molecular and a turbulent contribution ($\mu_t$), and the system is closed using a  two-equation turbulence model to determine $\mu_t$ and $k$. Details of the turbulence modelling practice are omitted to safe space and can be found in textbooks, e.g. \citep{wilcox1998turbulence}. 

%%%%%%%%%%%%%%%%%%%%%%%%%%%%%%%%%%%%%%%%%%%%%
\subsection{Equation of State}
\label{sec:c-state}
We expect both fluid phases to be incompressible and virtually immiscible. Moreover, we assume  no-slip between the fluid phases along the interface and model the flow as a mixture between fluids which share the velocity field governed by equation (\ref{equ:prima_momen_with_dimenX}). The continuity equation (\ref{equ:prima_mass_with_dimenX}) serves to determine the pressure and the  local fluid properties $\rho$ and $\mu$  follow from an equation of state. A more general equation of state (EoS) refers to a weighted sum of the bulk properties of the participating phases, viz. 
\begin{align}
\rho = m \rho_\mathrm{a} + \left[ 1 - m  \right] \rho_\mathrm{b}, \label{equ:mater_prope}
\qquad
\mu = m \mu_\mathrm{a} + \left[ 1 - m \right] \mu_\mathrm{b}.
\end{align}
The normalized function $m(c)$ describes the transition of the properties and only depends on the concentration $c$.The volume concentration of the foreground phase is physically bounded by $c \in [0,1]$. A frequently employed simple choice for the transition function refers to a linear interpolation
\begin{align}
m = \begin{cases}
0 &\text{if } \ c < 0 \\
c &\text{if } \ 0 \leq c \leq 1 \\
1 &\text{if } \ c > 1
\end{cases}   \label{equ:m-func1} \, . 
\end{align}
An alternative approach employed herein reads
\begin{align}
m = \frac{1}{2} \left[ \mathrm{tanh} \left( \frac{2c-1}{\gamma_\mathrm{m}}\right) +1 \right]    \label{equ:m-func2} \, . 
\end{align}
Here $\gamma_\mathrm{m}$ is a non-dimensional model parameter to adjust the transition regime. For $0.1 \leq \gamma_\mathrm{m} \leq 0.5$ the difference between (\ref{equ:m-func1}) and (\ref{equ:m-func2}) is limited to 0.009-3.597\% at the transition points $c=0$ and $c=1$, resulting in a slight offset of fluid properties. The formulation (\ref{equ:m-func2}) serves the regularization of unbounded concentration values and helps to sharpen partly blurred interfaces.
%
%%%%%%%%%%%%%%%%%%%%%%%%%%%%%%%%%%%%
\subsection{Diffusive Interface Model}
\label{sec:ch}
The subsection briefly summarizes the diffusive interface model for isothermal two-phase flows as suggested in a landmark paper by Cahn and Hilliard \cite{cahn1958free} and later elucidated by Jacqmin \cite{jacqmin1999calculation}. The present approach refers to a classical CH-model and is based upon the free energy $E$ of the interface $\Gamma$ between two isothermal fluid phases 
\begin{align}
E = \int_{\Gamma} e \, 
%\left( c, \frac{\partial c}{\partial x_k} \right) 
\mathrm{d} \Gamma = 
 \int_{\Gamma} \left[ 
C_\mathrm{1} b \left( c \right) + \frac{C_\mathrm{2} }{2} {\left|\frac{\partial c}{\partial x_\mathrm{k}}\right|}^2 \right] 
\mathrm{d} \Gamma \; .
 \label{equ:globa_free_energ}
\end{align}
The coefficients $C_\mathrm{1} \SI{}{[Pa]}$ and $C_\mathrm{2}\SI{}{[N]}$  can be determined from the interface thickness $\gamma$[m] and surface tension  $\sigma_\mathrm{a,b}\SI{}{[N/m]}$ between the two fluids, as outlined below. The foreground phase concentration $c$ represents a measure of phase and ranges from a foreground state ($c_\mathrm{a}$) to a background 
state ($c_\mathrm{b}$), i.e. $c_\mathrm{a} = 1$ and $c_\mathrm{b} = 0$. Mind that alternative CH-formulations exist, which employ the mass concentration or other energetic contributions. 

The first term of $e$ refers to the bulk  energy density and aims to separate the fluids.  The second term represents the gradient energy which widens the interface. To model separated (immiscible) fluids, a fourth-order polynomial, labelled  ''double well potential'', is often used to describe the bulk energy density, i.e. $b \left( c \right) = \left(c - c_\mathrm{a} \right)^2 \left( c - c_\mathrm{b} \right)^2$
\begin{align}
b \left( c \right) = \left(c - 1 \right)^2 c^2 .
\label{equ:CH-b}
\end{align}
In equilibrium conditions,  $E$ is minimized with respect to $c$. Using variational calculus, this relates to the root of a chemical potential $\psi$  for the  equilibrium state of plane interfaces (i.e. $\psi =0$)
\begin{align}
\psi = \frac{\delta E}{\delta c} = C_\mathrm{1}  \frac{\partial b}{\partial c}  - C_\mathrm{2}  \frac{\partial^2 c}{\partial {x_k}^2}  
% \label{equ:chemi_poten} 
\qquad \to \qquad 
C_\mathrm{1} \frac{\partial b}{\partial c} = C_\mathrm{2} \frac{\mathrm{d}^2 c}{\mathrm{d} {x_n}^2}, 
\label{equ:optim_situa}
\end{align}
where $x_\mathrm{n}$ represents the interface normal direction. Substituting (\ref{equ:CH-b}) into  (\ref{equ:optim_situa}), one obtains a hyperbolic tangent concentration profile. This also renders a relation between the coefficients $C_1, C_2$ and an interface thickness $\gamma$, viz.
\begin{align}
c \left( x_\mathrm{n} \right) = \frac{1}{2} \left( \mathrm{tanh} \left( \frac{2 x_\mathrm{n} }{\gamma} \right) + 1 \right) \label{equ:analy_solut} \qquad 
 {\rm with} \quad \gamma := \sqrt{\frac{8 C_\mathrm{2}}{C_\mathrm{1}}} \,  .
\end{align}
Similarly, surface tension forces can be related to the concentration. Jacqmin \cite{jacqmin1999calculation} outlined, that the convective rate of  change of the free energy widens or compresses the interface. The term reads $\delta E/\delta c \, (v_i \, \partial c/\partial x_i ) = \psi  \, ( v_i \, \partial c/\partial x_i )$ and should be balanced by the power $f_i^\text{ST} \, v_i$ of the surface tension force $f_i^\text{ST}$.  This immediately reveals the surface tension force used herein
\begin{align}
f_i^\text{ST}= \psi  \, \frac{\partial c}{\partial x_i}.
 \label{equ:free_energ_change} 
\end{align}
Some authors rearrange this definition into an apparent pressure term and a term involving the chemical potential gradient  \cite{song19}, i.e. $f_i^\text{ST}= \partial(\psi c)/\partial x_i - c  \, \partial \psi/\partial x_i$. Similarly, other authors employ the relation
\begin{align}
\psi  \, \frac{\partial c}{\partial x_i} - \left[ C_\mathrm{1} \frac{\partial b}{\partial x_i} 
+ \frac{C_2}{2}  \frac{\partial }{\partial x_i} \left( \frac{\partial c }{\partial x_k} \right)^2 \right] 
=
- \frac{ \partial }{\partial x_k} \left( C_2 \, \frac{\partial c }{\partial x_k} \frac{\partial c }{\partial x_i} \right)
 \label{equ:alt_sforce}
\end{align}
to express the surface tension force by the divergence of a surface tension stress $-C_2 \, (\nabla_k c)(\nabla_i c)$ and the gradient of a related apparent pressure $C_1 b + 0.5 \, C_2 (\nabla_k c)^2$ \cite{lowengrub1998quasi,abels2012thermodynamically}. Expression (\ref{equ:alt_sforce})  associates vanishing surface tension effects with $C_2=0$. Jacqmin \cite{jacqmin1999calculation} also deduced a link between the surface tension and $C_1$, $C_2$ for a plane interface 
\begin{align}
\sigma_\mathrm{a,b} = C_\mathrm{2} \int_{-\infty}^{\infty} \left( \frac{\partial c}{\partial x_\mathrm{n} } \right)^2 \mathrm{d} x_\mathrm{n}  \label{equ:surfa_tensi_coeff}.
\end{align}
Substituting  (\ref{equ:analy_solut}) into  (\ref{equ:surfa_tensi_coeff}) yields
\begin{align}
\sigma_\mathrm{a,b} = \sqrt{ \frac{C_\mathrm{1} C_\mathrm{2}}{18}} \label{equ:surf_tensi_final}.
\end{align}
Once $\gamma$ and $\sigma_\mathrm{a,b}$ have been chosen, both CH coefficients can be determined from the plane interface relations (\ref{equ:analy_solut}, \ref{equ:surf_tensi_final})
\begin{align}
C_\mathrm{1}  = 12 \frac{\sigma_\mathrm{a,b}}{\gamma} \, ,  \qquad
C_\mathrm{2}  = \frac{3}{2} \sigma_\mathrm{a,b} \gamma.
\end{align}
In the remainder of the paper,  the neglect of the surface tension in under-resolved flows is modelled by $C_2=\SI{0}{N}$ and, for the sake of simplicity, $C_1= \SI{1}{Pa}$. 

%%%%%%%%%%%%%%%%%%%%%%%%%%%%%%%%%%%%%%%%%%%%
\subsubsection{Concentration Transport and Velocity Divergence}
\label{sec:c-transport}
The mixture fraction is computed from an additional transport equation that models the mass transfer between the phases. Two options will be discussed, referring to the classical VoF and the CH approach outlined in Section \ref{sec:ch}. Using the classical VoF approach, we assume that the material property of the fluid must not change, viz.
\begin{align}
\frac{\mathrm{d} c}{\mathrm{d} t} = \frac{\partial c}{\partial t} +v_\mathrm{i} \frac{\partial  c}{\partial x_\mathrm{i}} = 0 \, .  \label{equ:VoF}
\end{align}
On the contrary, the CH-approach of Lowengrub and Truskinovsky  \cite{lowengrub1998quasi} refers to the mass concentration $c_m= c \rho_1/\rho$ of the foreground phase and involves a diffusive phase transfer term
\begin{align}
\rho \frac{\mathrm{d} c_m}{\mathrm{d} t} =  \frac{\partial }{\partial x_\mathrm{i}} \left( M_m \frac{\partial  \psi}{\partial x_\mathrm{i}} 
\right) \, . \label{equ:CH1x}
\end{align}
Using the continuity relation (\ref{equ:prima_mass_with_dimenX}) leads to 
\begin{align}
 \frac{\partial \, \rho c_m}{\partial t} + \frac{\partial  \, \rho v_i c_m}{\partial x_\mathrm{i}} = \frac{\partial }{\partial x_\mathrm{i}} 
\left( M_m \frac{\partial  \psi}{\partial x_\mathrm{i}} \right) \, .
\label{equ:CH1b}
\end{align}
The desired conservative volume-flux based transport follows from (\ref{equ:CH1b}), with $\rho c_m = \rho_1 c$ and $M_m=M/\rho_1$ 
\begin{align}
 \frac{\partial c}{\partial t} + \frac{\partial  \, v_i c}{\partial x_\mathrm{i}} 
= \frac{\partial }{\partial x_\mathrm{i}} 
\left( M \frac{\partial  \psi}{\partial x_\mathrm{i}} \right) \, .
\label{equ:CH1a}
\end{align}
Here the mobility parameter $M$ of dimension $\SI{}{m^3  s / kg}$ $ \left[\hat{=} \ \SI{}{\nu / Pa} \right]$ is a free parameter that controls the strength of the phase separation process and will be explored for under-resolved flows in Section \ref{sec:appro_mobil_estim}.
%%%%%%%%%%%%%%%%%%%%%%%%%%%%%%%%%%%5
%\subsubsection{Continuity Condition}
%\label{sec:incompcond}

VoF methods for immiscible incompressible fluids are usually based upon a pressure correction/projection scheme and prefer to observe volume instead of the mass fluxes to avoid the density jump. The latter can be extended for hydrodynamic flows featuring $\partial v_\mathrm{i} / \partial x_\mathrm{i} = 0$ along the route outlined in \citep{yakubov2015experience}.
Since the bulk densities are deemed incompressible, the density solely depends on the transition function $m$, which in turn depends on the concentration (\ref{equ:m-func1}), (\ref{equ:m-func2}). An alternative continuity equation  is derived from  (\ref{equ:prima_mass_with_dimenX})
\begin{align}
 \left( \frac{\rho_b - \rho_a }{\rho}  \right) \, \frac{\partial m}{\partial c} \; \left[ \frac{\partial c}{\partial t} +v_i \frac{\partial  c}{\partial x_\mathrm{i}} \right] = \frac{\partial v_\mathrm{i}}{\partial x_\mathrm{i}}   \, .
 \label{equ:MIX1}
\end{align}
Substituting  (\ref{equ:CH1a}) into  (\ref{equ:MIX1}), one finally arrives at
\begin{align}
  \frac{\partial v_i}{\partial x_i}  = f \,   \frac{\partial }{\partial x_\mathrm{i}} \left( M \frac{\partial  \psi}{\partial x_\mathrm{i}} \right)  \qquad
 {\rm with} \quad  f=\frac{f^*}{1+f^*c}  \quad {\rm and} \quad f^*= \, 
\left(\frac{\rho_b - \rho_a}{\rho} \right) \frac{\partial m}{\partial c} \label{equ:MIX2a} .
\end{align}
In conjunction with VoF, a solenoidal velocity field is recovered due to the neglect of diffusive mass transfer. Note that $f^*$  is virtually zero in combination with the non-linear EoS (\ref{equ:m-func2}) in the sharp interface limit, which returns a divergence-free velocity field.

%%%%%%%%%%%%%%%%%%%%%%%%%%%%%%%%%%%%%%%%%%%
\subsection{Non-dimensional Governing Equations}
\label{sec:nondim}
The non-dimensional  equations support the discussion of  influences and assist the verification. Assuming a spatially constant mobility $M$, the  non-conservative concentration, continuity and momentum equations read:
\begin{alignat}{2}
& \frac{1}{\mathrm{St}} \frac{\partial c}{\partial t^*} +  v_\mathrm{i}^* \frac{\partial \;  c }{\partial x_\mathrm{i}^*}  = \frac{1}{\mathrm{Pe}} \frac{\partial^2 }{\partial x_\mathrm{i}^{*2}} \left[ 12 \frac{\partial b^*}{\partial c}  - \frac{3 \, Ca^2}{2}  \frac{\partial^2 c}{\partial x_\mathrm{k}^{*2}} \right]   \, \left(\frac{1}{1 + f^*c} \right)  && \; =  \frac{1}{\mathrm{Pe}} \, \left(\frac{1}{1 + f^*c} \right)  \;  \frac{\partial^2 \psi^* }{\partial x_\mathrm{i}^{*2}}   \label{equ:prima_conce_dile} \\
&\frac{\partial v_\mathrm{i}^*}{\partial x_\mathrm{i}^*}  && \; =  \frac{f}{\mathrm{Pe}}  \; \frac{\partial^2 \psi^* }{\partial x_\mathrm{i}^{*2}} \label{equ:prima_mass_dile} \\
&\frac{\rho^*}{\mathrm{St}}  \frac{\partial v_\mathrm{i}^*}{\partial t^*} + \rho^* v_\mathrm{j}^* \frac{\partial v_\mathrm{i}^*}{\partial x_\mathrm{j}^*} + \frac{\partial }{\partial x_\mathrm{j}^*} \left[\mathrm{Eu} \  p^* \delta_\mathrm{ij} - \frac{2 \mu^*}{\mathrm{Re}}  S_\mathrm{ij}^* \right] -  \frac{\rho^*}{\mathrm{Fn}^2}  g_\mathrm{i}^* && \; = -\frac{2}{3} \frac{1}{\mathrm{Re}}  \frac{\partial}{\partial x_i^*} \left[ \mu^* \frac{f}{\mathrm{Pe}}  \; \frac{\partial^2 \psi^* }{\partial x_\mathrm{i}^{*2}} \right]  + \frac{\psi^*}{\mathrm{We}}  \; \frac{\partial c}{\partial x_\mathrm{i}^*} \label{equ:prima_momen_dile}
\end{alignat}
The left-hand sides of the balance equations represent the classical, incompressible VoF-method. In the case of a non-zero right-hand side of (\ref{equ:prima_conce_dile}), the velocity field is no longer divergence free in (\ref{equ:prima_mass_dile}), which in turn introduces additional terms to the momentum balance  (\ref{equ:prima_momen_dile}). 
An exemplary relationship between a dimensional quantity, a  reference value and a non-dimensional quantity marked with an asterisk reads $v_\mathrm{i} = V v_\mathrm{i}^*$. The dimensionless parameters are defined by
\begin{alignat}{4} \label{equ:dimen_numbe}
\mathrm{St}  &= \frac{\mathrm{T} \ \mathrm{V}}{\mathrm{L}} \ (\mathrm{Strouhal}) \hspace{5mm} &&\mathrm{Eu} = \frac{\mathrm{P}}{ \rho_\mathrm{b} \ \mathrm{V}^{2} } \ (\mathrm{Euler}) \hspace{5mm} &&\mathrm{Re}  =  \frac{\rho_\mathrm{b} \ \mathrm{V} \ \mathrm{L}}{\mu_\mathrm{b}} \ (\mathrm{Reynolds})  \hspace{5mm} &&\mathrm{We}  =  \frac{ \rho_\mathrm{b} {V}^2 \gamma}{\sigma_\mathrm{a,b}} \ (\mathrm{Weber}) \nonumber \\
\mathrm{Fn}  &= \frac{\mathrm{V}}{ \sqrt{ G \ \mathrm{L} }} \ (\mathrm{Froude}) &&\mathrm{Pe}  = \frac{\mathrm{V} \ \mathrm{L} \gamma}{ \mathrm{M} \sigma_\mathrm{a,b}} \ (\mathrm{Peclet}) &&\mathrm{Ca} = \frac{\mathrm{\gamma}}{L} \ (\mathrm{Cahn}).
\end{alignat}
The quantities utilized for the non-dimensionalisation are given in Table \ref{tab:refe_data}. Local discrete similarity parameters employ $L=\delta x_i$, $T=\delta t$ and $V= \| v_i\|$ etc.. It should be pointed out that the transition length $\gamma $ can be small compared to a  local grid spacing $\delta x_i$, resulting in small (discrete) Cahn-numbers, which supports the neglect of the second term on the r.h.s. of (\ref{equ:prima_conce_dile}) in an under-resolved  sharp interface limit.

\begin{table}[h]
\centering
\begin{tabular}{|c||c|c|c|c|c|c|c|c|c|}
\hline
field quantity & 
$x_\mathrm{i}$  & $v_\mathrm{i}$ & $g_\mathrm{i}$ & $p$ & $t$ & $\rho$ & $\mu$ & $c$ & $\psi$ \\
\hline
reference value & $\mathrm{L}$ & $\mathrm{V}$ & $\mathrm{G}$ & $\mathrm{P}$ & $\mathrm{T}$ & $\rho_\mathrm{b}$ & $\mu_\mathrm{b}$ & 1 & $\sigma_{a,b}/\gamma$\\
\hline
\end{tabular}
\caption{Quantities for the non-dimensionalisation of the governing equations.}
\label{tab:refe_data}
\end{table}

%%%%%%%%%%%%%%%%%%%%%%%%%%%%%%%%%%%%%%%%%%%
\section{Numerical Procedure}
\label{sec:numer_proce}
Results of the present study are obtained from the Navier-Stokes procedure FreSCo$^+$ \citep{rung2009challenges}. The implicit finite volume procedure uses a segregated algorithm based on the strong conservation form and employs a cell-centred, co-located storage arrangement for all transport properties. Unstructured grids, based on arbitrary polyhedral cells or hanging nodes, can be used. 
The solution is iterated to convergence using a modified pressure-correction scheme \citep{yakubov2015experience}. Various turbulence-closure models are available with respect to statistical (RANS) or scale-resolving (LES, DES) approaches. Time derivatives are approximated by an implicit Euler or implicit three time level (ITTL) scheme. The numerical integration employs a second-order mid-point rule, diffusive fluxes are determined from second-order accurate central differencing and convective fluxes use higher-order upwind biased interpolation formulae.
Since the data structure is generally unstructured, suitable preconditioned iterative sparse-matrix solvers for symmetric and non-symmetric systems, e.g. GMRES, BiCG, QMR, CGS or BiCGStab, are used. The procedure is parallelized for several thousand processes using a domain decomposition methods and the MPI communications protocol \citep{yakubov2013hybrid}. It supports local mesh refinement, overset grids \citep{volkner2017analysis}, node-based adjoint shape-optimization \cite{kroger2015cad, kroger2018adjoint} and fluid-structure interactions between mechanically coupled floating bodies \citep{luo2017computation}.

Cell-centred fluid properties are determined from (\ref{equ:mater_prope}). For face-based properties, a linear interpolation between adjacent cell-centre values is used.  The discretization of the additional terms that originate from the CH-NS approach are discussed in the upcoming lines for equations (\ref{equ:CH1a} - concentration), (\ref{equ:MIX2a}  - continuity) and (\ref{equ:prima_momen_with_dimenX}  - momentum). The discussion refers to the symbolic finite-volume approximation of a variable $\phi$ located in the center $P$ of  a control volume  of size $\delta \Omega_P$, with neighbour control columns $NB$, i.e. $A_P \phi_P - \sum_{NB} A_{NB} \phi_{NB} = S_{\phi} \, \delta \Omega_P$. Here, the right-hand side source term $S_{\phi}$ is treated explicitly during the (time-implicit) iteration of a segregated solution procedure \cite{ferziger2012computational}.
All presented own VoF studies employ the compressive HRIC scheme for convective concentration transport \cite{muzaferija1998computation}.

\subsection{Concentration Conservation}
\label{sec:num-c}
The numerical solution of (\ref{equ:CH1a}) follows a deferred correction approach. Hereto the right-hand side  is notionally split into a bulk density contribution and a  gradient term, i.e. $\nabla_i [M  (\nabla_i \psi )]=\nabla_i [M  (\nabla_i (  C_1  \partial b/\partial c) ] + \nabla_i [\dots] $. Using $\nabla_i b = (\partial b/\partial c) \nabla_i c$ and $\partial^2 b/\partial c^2 = 2 + 12c \, (c-1)$, an   inherently positive contribution to the bulk density term is identified. This leads to an implicit contribution and an explicit source $S_c$ which uses values of the previous iteration
\begin{align}
\frac{\partial} {\partial {x_\mathrm{i}}} \left( M \frac{\partial \psi}{\partial {x_\mathrm{i}}} \right) = 
\frac{\partial}{\partial x_\mathrm{i}} \left(  2 M \, C_1 \frac{\partial c}{\partial x_\mathrm{i}} \right)
+
\underbrace{\left[ 
\frac{\partial} {\partial {x_\mathrm{i}}} \left( M \frac{\partial \psi}{\partial {x_\mathrm{i}}} \right) 
- \frac{\partial}{\partial x_\mathrm{i}} \left(  2 M \, C_1 \frac{\partial c}{\partial x_\mathrm{i}} \right)
\right]}_{S_c}
% 
% M \frac{\partial^2}{\partial x_\mathrm{i}^{2}} \left[ C_1 \frac{\partial b}{\partial c} \right] = 
%\frac{\partial}{\partial x_\mathrm{i}}\left[  \left( M \, C_1 \frac{\partial^2 b}{\partial c^2}  \right) \frac{\partial c}{\partial x_\mathrm{i}} \right]
% = 
%\frac{\partial}{\partial x_\mathrm{i}} \left[  2 M \, C_1 \frac{\partial c}{\partial x_\mathrm{i}} \right]
%+ 
%\; 
% \underbrace{\frac{\partial^2}{\partial x_\mathrm{i}^2} \, \left[ M \, C_1 \, (4c^3 -6c^2) \right]}_{S_{c(2)}}
\; . 
\label{equ:cahn_hilli_diffux}
\end{align}
The integration over a control volume yields a discretized  surface integral over all sub surfaces, i.e. $\delta \Gamma_{i }^{(f)}$ of $\delta \Omega_P$ for the implicit part of (\ref{equ:cahn_hilli_diffux}) $\sum_{f(\delta \Omega_P)}   (2 M \, C_1) \nabla_i  c  \, \delta \Gamma_{i}$, which is discretized using central differences and the mid-point integration rule. The  explicit  source term  follows  from  a mid-point integration over the control volume. Upwind biased schemes are used to approximate the convective fluxes of (\ref{equ:CH1a}) for the CH approach and a compressive downwind biased approach \cite{muzaferija1998computation} is used for the VoF approach. 

\subsection{Momentum and Mass Conservation}
A simple mid point integration is employed to account for the additional explicit CH-NS sources. This involves  $f \nabla_i (M \nabla_i \psi)   \delta \Omega_P$ for the conservation of mass (\ref{equ:MIX2a}) in combination with a pressure correction scheme \citep{yakubov2015experience}, and $f_i^{\rm ST} \delta \Omega_P$ (\ref{equ:free_energ_change}) for the momentum equation (\ref{equ:prima_momen_with_dimenX}). 

%%%%%%%%%%%%%%%%%%%%%%%%%%%%%%%%%%%%%%%%%%%
\section{Verfication}
\label{sec:one_dimen}

\subsection{Planar Couette Flow}
\begin{figure}[h]
	\centering
	\subfigure[]{
		\begin{tikzpicture}
\filldraw[pattern=north east lines, pattern color=black] (0,0) rectangle (7,0.25);
\filldraw[fill=mycolor_blue!100, draw=none] (0,0.25) rectangle (7,2);
\filldraw[fill=mycolor_grey!100, draw=none] (0,2.0) rectangle (7,4);
\draw[thin] (0,0) -- (7,0);
\draw[thin] (0,0.25) -- (7,0.25);
\filldraw[pattern=north east lines, pattern color=black] (0,4) rectangle (7,4.25);
\draw[thin] (0,4) -- (7,4);
\draw[thin] (0,4.25) -- (7,4.25);
\draw[thin,->] (6,3.75) -- (7,3.75) node[anchor=west] {$v_\mathrm{top}$};
\draw[dotted] (0,2.0) -- (7,2.0);
\draw (4.5,0.8) node[] {$c = 1$};
\draw (4.5,2.5) node[] {$c = 0$};
\draw[thin,->] (0,0.25) -- (8,0.25) node[anchor=south] {$x_\mathrm{1}$};
\draw[thin,->] (0.25,0) -- (0.25,1.5) node[anchor=south] {$x_\mathrm{2}$};

\draw[thin,->] (7.5,2.5) -- (7.5,1.25) node[anchor=west] {$g$};
%\draw[thin,->] (7.5,2.5) -- (8.25,1.25) node[anchor=west] {$g$};
%\draw[dashed] (7.5,2.5) -- (7.5,1.25);
%\draw[dashed] (7.5,1.25) -- (8.25,1.25);
%\draw [thin] (7.5,1.75) to [out=0,in=250] (7.8,2.0);
%\draw (7.625,2.0) node[] {$\varphi$};

\draw[thin,->] (1.,0.25) -- (1.,2.0);
\draw (1.0,1.25) node[anchor=west] {$h_\mathrm{m}$};
\draw[thin,->] (2,0.25) -- (2,3.95);
\draw (1.5,3.75) node[anchor=west] {h};
\draw (4.5,3.75) node[] {$p_\mathrm{top}$};
\end{tikzpicture}
	}
	\hspace{2cm}
	\subfigure[]{
		\includegraphics[scale=0.595]{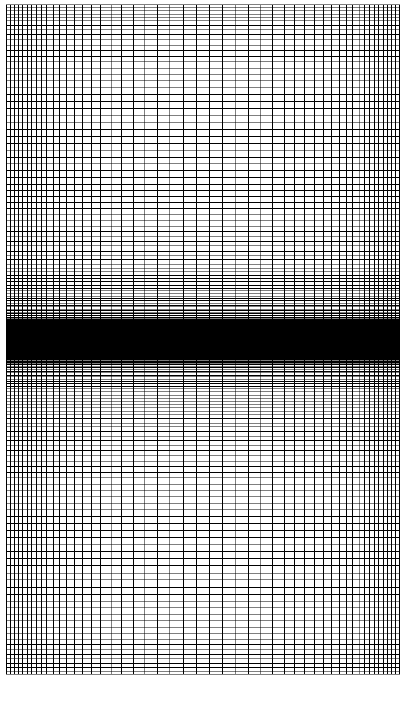}
	}
	\caption{(a) Setup of the planar Couette flow verification case  and (b) employed computational grid.}
	\label{fig:couet_flow}
\end{figure}
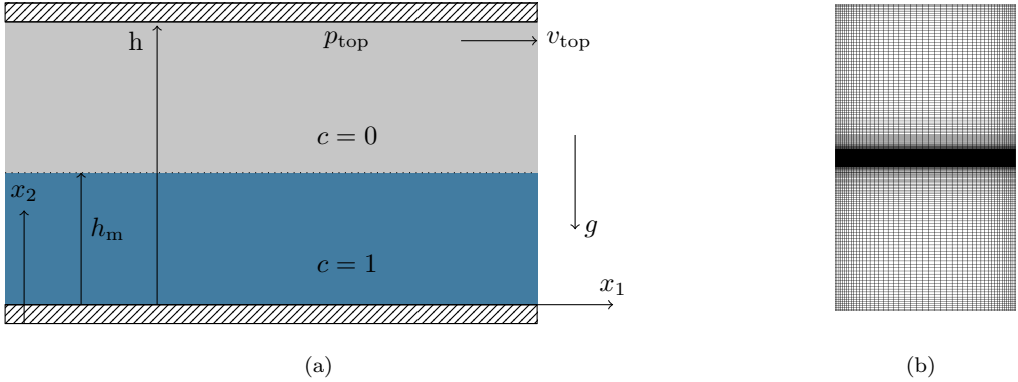

The implementation is verified for a planar Couette flow under the influence of vertical gravity, wherefore a non-dimensional analytical solution is constructed and compared with the numerical results. Figure \ref{fig:couet_flow} illustrates the considered test case and the employed numerical grid which involves $50 \times 600$ control volumes. The channel height refers to $h$.
The lower half of the channel is filled with a dense background fluid and the free surface follows along a horizontal line $x_2=h/2$, where the origin of the coordinate system is located at the bottom wall. The bottom wall is at rest and the top wall moves with  $v_1 = v_{top}$ along the horizontal direction. Field values are non-dimensionalised with the reference quantities, $V=v_{top}$, $L=h$ and $P=\rho_\mathrm{b} ||g_\mathrm{i}|| h$ as well as the non-dimensional equation system (\ref{equ:prima_conce_dile}-\ref{equ:prima_momen_dile}) is used.
 
The velocity is assumed to be unidirectional, i.e. $v_1^*(x_2^*)$, and in a  fully developed, laminar, steady state. Moreover, the concentration field is also considered steady and  homogeneous in the primary direction ($x_1^*$), viz. 
\begin{alignat}{2}
\mathrm{C}&:  \frac{\partial^2 }{\partial {x_2^*}^2} \left[ 4c^3 - 6c^2 + 2c - \mathrm{Ca}^2 \frac{\partial^2 c}{\partial {x_2^*}^2} \right] &&= 0 \, , \label{equ:finit_trans_conc_simpl} \\
\mathrm{R}_1 &:  \frac{\partial }{\partial x_2^*} \left[ \mu^* \frac{\partial v_1^*}{\partial {x_2^*}} \right] &&= 0 \, ,  \label{equ:finit_trans_r1_simpl} \\
\mathrm{R}_2 &:\mathrm{Eu} \frac{\partial p^*}{\partial x_2^*} - \frac{1}{\mathrm{Fn}^2} \rho^* g_2^* &&= 0  \, .  \label{equ:finit_trans_r2_simpl}
\end{alignat}
Using the linear EoS (\ref{equ:m-func1}) allows an integration of (\ref{equ:finit_trans_conc_simpl}-\ref{equ:finit_trans_r2_simpl}) and results in the following analytical solution
\begin{align}
c &= \frac{1}{2} \left[ \mathrm{tanh} \left( \mathrm{Ca}^{-1}  \left(2x_2^*-1 \right) \right) + 1 \right] \label{equ:finit_trans_conce_solut}  \, ,  \\
v_1^* &= \frac{ 4 \ x_\mathrm{2}^* +  \mathrm{Ca}  \, \left( \mu_\mathrm{a}^* - 1 \right) \mathrm{log} \left[ \frac{\mu_\mathrm{a}^* + 1 + \left( 1 - \mu_\mathrm{a}^* \right) \ \mathrm{tanh}\left( \mathrm{Ca}^{-1} \right)}{\mu_\mathrm{a}^* + 1 - \left( 1 - \mu_\mathrm{a}^* \right) \ \mathrm{tanh}\left( \mathrm{Ca}^{-1} \left( 2 x_\mathrm{2}^* - 1 \right) \right)} \frac{\mathrm{tanh} \left( \mathrm{Ca}^{-1} \left( 2 x_\mathrm{2}^* - 1 \right) \right) + 1}{1 - \mathrm{tanh} \left( \mathrm{Ca}^{-1} \right)} \right] }{ 4  +\mathrm{Ca} \  \left( \mu_\mathrm{a}^* - 1 \right) \mathrm{log} \left[ \frac{\mu_\mathrm{a}^* + 1 + \left( 1 - \mu_\mathrm{a}^* \right) \ \mathrm{tanh}\left( \mathrm{Ca}^{-1} \right)}{\mu_\mathrm{a}^* + 1 - \left( 1 - \mu_\mathrm{a}^* \right) \ \mathrm{tanh}\left( \mathrm{Ca}^{-1} \right)} \frac{\mathrm{tanh} \left( \mathrm{Ca}^{-1} \right) + 1}{1 - \mathrm{tanh} \left( \mathrm{Ca}^{-1} \right)} \right]  } \label{equ:finit_trans_veloc_solut}  \, , \\
p^* &= \frac{1}{\mathrm{Eu} \ \mathrm{Fn}^2} \left[  \rho_\mathrm{a}^* \left( 1 - x_\mathrm{2}^* \right) + \frac{\mathrm{Ca}  \, \rho_\Delta^*}{4 \ } \mathrm{log} \left( \frac{\mathrm{tanh}\left( \mathrm{Ca}^{-1}  \left( x_\mathrm{2}^* - 1 \right) \right) + 1}{\mathrm{tanh} \left( \mathrm{Ca}^{-1} \right) + 1} \right) \right] . \label{equ:finit_trans_press_solut} 
\end{align}
This solution follows from Dirichlet conditions for the concentration and velocities along the top as well as the bottom wall ($v_1^*(1)=c(0)=1, \; v_1^*(0)=c(1)=0$). 
Additionally, a prescribed top-wall pressure ($p^*(1)=0$) is employed.
% . 
Interestingly, the solution is independent from the Peclet number und thus also from the mobility parameter $M$.  

Computational results for the CH-NS system (\ref{equ:prima_momen_with_dimenX},\ref{equ:CH1a},\ref{equ:MIX2a}) are obtained on a 2D grid in conjunction with periodic boundary conditions in stream wise direction. Convective fluxes for momentum and concentration are discretized using first-order upwind differencing (UDS) and the mobility is assigned to a value that results in $\mathrm{Pe}  = \SI{1E+05}{}$.  
Predictive results are compared with analytical solutions for a range of  Cahn numbers (Ca$_h$=$0.2, 0.1, 0.02$), viscosity ratios ($\mu_a^*=0.25, 1, 4$) and density ratios ($\rho_a^*=0.25, 1, 4$) along with exemplary flow conditions of $\mathrm{Re} = \SI{100}{}$, $\mathrm{Fn} = \SI{0.5}{}$ and $\mathrm{Eu} = 4$.  The vertical interface resolution involves 20, 100 or 200 control volumes depending on the Cahn number. 
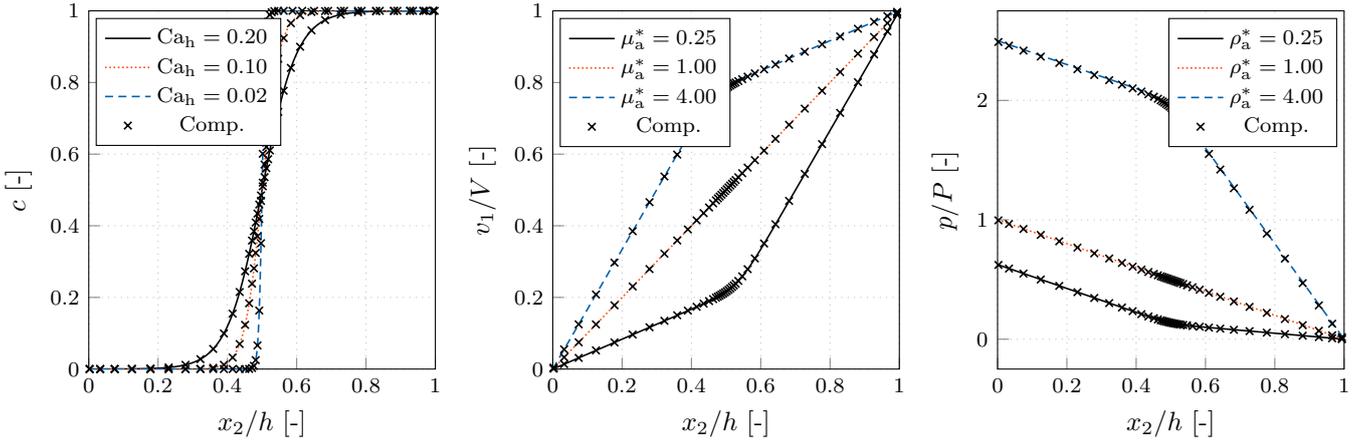
\begin{figure}[h]
\centering
\analytiSolutionPictures
\begin{tikzpicture}
\begin{axis}[
 xlabel style={text width=0.25\textwidth,align=center},
 ylabel style={text width=0.25\textwidth,align=center},
 xlabel={$x_\mathrm{2} / h$ [-]},
 ylabel={$c$ [-]},
 xmin=0,xmax=1,
 legend style={at={(0.02,0.98)},anchor=north west},
 ymin=0.0,ymax=1.0
]
\addplot [line1] table[x expr={\thisrowno{0}},y expr={\thisrowno{1}}]{data/Couette_Concentration_Analytic_Results.dat};
\addplot [line2] table[x expr={\thisrowno{0}},y expr={\thisrowno{2}}]{data/Couette_Concentration_Analytic_Results.dat};
\addplot [line3] table[x expr={\thisrowno{0}},y expr={\thisrowno{3}}]{data/Couette_Concentration_Analytic_Results.dat};
\addplot [mark1, only marks,mark repeat=5] table[x expr={\thisrowno{0}},y expr={\thisrowno{1}}]{data/Couette_Concentration_FreSCo_CaC_5_Results.dat};
\addplot [mark1, only marks,mark repeat=5] table[x expr={\thisrowno{0}},y expr={\thisrowno{1}}]{data/Couette_Concentration_FreSCo_CaC_10_Results.dat};
\addplot [mark1, only marks,mark repeat=5] table[x expr={\thisrowno{0}},y expr={\thisrowno{1}}]{data/Couette_Concentration_FreSCo_CaC_50_Results.dat};
 
\addlegendentry{$\mathrm{Ca}_\mathrm{h} = 0.20$};
\addlegendentry{$\mathrm{Ca}_\mathrm{h} = 0.10$};
\addlegendentry{$\mathrm{Ca}_\mathrm{h} = 0.02$};
\addlegendentry{Comp.};
 
\end{axis}
\end{tikzpicture}
\begin{tikzpicture}
\begin{axis}[
 ylabel style={text width=0.25\textwidth,align=center},
 xlabel={$x_\mathrm{2} / h$ [-]},
 ylabel={$v_\mathrm{1} / V$ [-]},
 xmin=0,xmax=1,
 legend style={at={(0.02,0.98)},anchor=north west},
 ymin=0.0,ymax=1.0
]
\addplot [line1] table[x expr={\thisrowno{0}},y expr={\thisrowno{1}}]{data/Couette_Velocity_Analytic_Results.dat};
\addplot [line2] table[x expr={\thisrowno{0}},y expr={\thisrowno{2}}]{data/Couette_Velocity_Analytic_Results.dat};
\addplot [line3] table[x expr={\thisrowno{0}},y expr={\thisrowno{3}}]{data/Couette_Velocity_Analytic_Results.dat};
\addplot [mark1, only marks,mark repeat=5] table[x expr={\thisrowno{0}},y expr={\thisrowno{1}}]{data/Couette_Velocity_FreSCo_muD_m_025_Results.dat};
\addplot [mark1, only marks,mark repeat=5] table[x expr={\thisrowno{0}},y expr={\thisrowno{1}}]{data/Couette_Velocity_FreSCo_muD_0_Results.dat};
\addplot [mark1, only marks,mark repeat=5] table[x expr={\thisrowno{0}},y expr={\thisrowno{1}}]{data/Couette_Velocity_FreSCo_muD_p_4_Results.dat};
 
\addlegendentry{$\mu_\mathrm{a}^* = 0.25$};
\addlegendentry{$\mu_\mathrm{a}^* = 1.00$};
\addlegendentry{$\mu_\mathrm{a}^* = 4.00$};
\addlegendentry{Comp.};

\end{axis}
\end{tikzpicture}
\begin{tikzpicture}
\begin{axis}[
 ylabel style={text width=0.25\textwidth,align=center},
 xlabel={$x_\mathrm{2} / h$ [-]},
 ylabel={$p / P$ [-]},
 xmin=0,xmax=1,
]
\addplot [line1] table[x expr={\thisrowno{0}},y expr={\thisrowno{1}}]{data/Couette_Pressure_Analytic_Results.dat};
\addplot [line2] table[x expr={\thisrowno{0}},y expr={\thisrowno{2}}]{data/Couette_Pressure_Analytic_Results.dat};
\addplot [line3] table[x expr={\thisrowno{0}},y expr={\thisrowno{3}}]{data/Couette_Pressure_Analytic_Results.dat};
\addplot [mark1, only marks,mark repeat=5] table[x expr={\thisrowno{0}},y expr={\thisrowno{1}}]{data/Couette_Pressure_FreSCo_rhoD_m_05_Results.dat};
\addplot [mark1, only marks,mark repeat=5] table[x expr={\thisrowno{0}},y expr={\thisrowno{1}}]{data/Couette_Pressue_FreSCo_rhoD_0_Results.dat};
\addplot [mark1, only marks,mark repeat=5] table[x expr={\thisrowno{0}},y expr={\thisrowno{1}}]{data/Couette_Pressure_FreSCo_rhoD_p_05_Results.dat};

\addlegendentry{$\rho_\mathrm{a}^* = 0.25$};
\addlegendentry{$\rho_\mathrm{a}^* = 1.00$};
\addlegendentry{$\rho_\mathrm{a}^* = 4.00$};
\addlegendentry{Comp.};

\end{axis}
\end{tikzpicture}
\caption{Comparison of numerical and analytical results for the planar Couette flow example at $\mathrm{Re} = 100$, $\mathrm{Fn} = \SI{0.5}{}$, $\mathrm{Eu} = 4$ and $\mathrm{Pe} = \SI{1E+05}{}$. Left: Concentration profiles for different Cahn-numbers; Middle: Velocity profiles for different viscosity ratios (Ca$_\mathrm{h}$=$0.1$); Right: Pressure profiles for different density ratios (Ca$_\mathrm{h}$=$0.1$).}
\label{fig:prima_resul_finit_TL}
\end{figure}

Numerical results extracted along the centre vertical line are displayed in  Figure \ref{fig:prima_resul_finit_TL}. The left graph compares analytical and computed concentration profiles for three different Cahn-numbers. A comparison of results obtained for different fluid properties at Ca$_h$=$0.1$  is displayed in the other two graphs of the figure. All comparisons reveal an excellent predictive agreement with the analytical solutions (\ref{equ:finit_trans_conce_solut})-(\ref{equ:finit_trans_press_solut}).

\subsection{Stationary Bubble}
\begin{figure}[h]
\subfigure[]{
\input{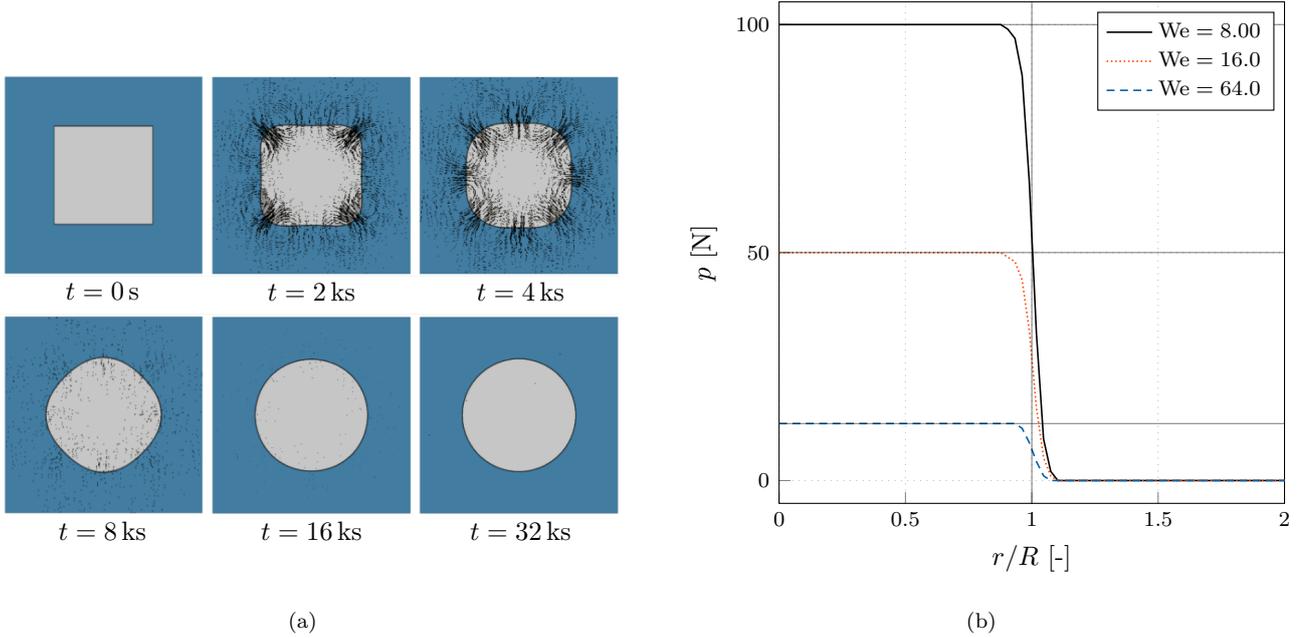}
}
\subfigure[]{
\smallPicture
\begin{tikzpicture}
\begin{axis}[
 ylabel style={text width=0.25\textwidth,align=center},
 xlabel={$ r/R $ [-]},
 ylabel={$p$ [N]},
 xmin=0.0,xmax=2.0,
 ymin=-5.0,ymax=105.0
]
\addplot [line1] table[x expr={\thisrowno{0}},y expr={\thisrowno{1}}]{data/Stationary_Bubble_sigma_5E-01.dat};
\addplot [line2] table[x expr={\thisrowno{0}},y expr={\thisrowno{1}}]{data/Stationary_Bubble_sigma_2.5E-01.dat};
\addplot [line3] table[x expr={\thisrowno{0}},y expr={\thisrowno{1}}]{data/Stationary_Bubble_sigma_6.25E-02.dat};

\addplot [color7,opacity=0.5] table[x expr={\thisrowno{0}},y expr={\thisrowno{1}}]{data/Stationary_Bubble_Expected.dat};
\addplot [color7,opacity=0.5] table[x expr={\thisrowno{0}},y expr={\thisrowno{2}}]{data/Stationary_Bubble_Expected.dat};
\addplot [color7,opacity=0.5] table[x expr={\thisrowno{0}},y expr={\thisrowno{3}}]{data/Stationary_Bubble_Expected.dat};
\addplot [color7,opacity=0.5] table[x expr={\thisrowno{4}},y expr={\thisrowno{5}}]{data/Stationary_Bubble_Expected.dat};
 
\addlegendentry{$\mathrm{We} = 8.00$};
\addlegendentry{$\mathrm{We} = 16.0$};
\addlegendentry{$\mathrm{We} = 64.0$};

\end{axis}
\end{tikzpicture}
}
\caption{(a) Evolution of  shape and velocity fields for the surface tension driven transition from an initially rectangular to a round bubble ($ks$ refers to $10^3$ seconds) and (b) computed pressure distribution along a radial slice that originates in the center of a bubble with radius $R=L/\sqrt{\pi}$ for different surface tension coefficients $\sigma_\mathrm{a,b} = 0.5, 0.25,0.0625$ [N/m] at $\mathrm{Ca}_\mathrm{L} = 0.1$, $\mathrm{Pe}_\mathrm{L} = 200  [N/m] / \sigma_\mathrm{a,b}$ and $\mathrm{We} = 4  [N/m] / \sigma_\mathrm{a,b}$. Grey horizontal lines in (b) indicate the expected interior pressure obtained by the Young-Laplace law.}
\label{fig:stati_bubbl}
\end{figure}

The influence of the surface tension model is verified by computing the transition from an initial non-equilibrium (rectangular) bubble into an equilibrium (circular) bubble.  The example is restricted to advancing a 2D flow field without gravitational effects ($\mathrm{Fn} = \infty$) in pseudo time.
As outlined in Figure \ref{fig:stati_bubbl}, a lighter foreground phase ($\rho_\mathrm{a}/\rho_\mathrm{b}$=1/100) rectangle with an edge length of $L=\SI{0.005}{m}$ is initially embedded into a heavier phase, such that the surface tension directs the shape of the interface towards a circle. Equal viscosities  are employed for both fluids. Due to the symmetrical arrangement, only one quarter of the bubble is simulated on a  homogeneous isotropic grid. Symmetry conditions are placed along the two main axes and the outer boundaries of the domain. 
The grid employs $550 \times 550$ control volumes to cover the quartered domain of $L \times L$.
The Cahn number based on the initial edge length $L$ reads $\mathrm{Ca}_\mathrm{L} = 0.02$ and the transition  is resolved by 11 control volumes. In line with an assumed unit Reynolds number, we define the reference velocity as $V = \mu_\mathrm{b} / ( \rho_\mathrm{b} \ L)$. The mobility and the surface tension are chosen to end up with Peclet and Weber numbers $\mathrm{Pe}_\mathrm{L} = \SI{4E+04}{}$ and $\mathrm{We} = 800$ respectively.

As observed in Fig. \ref{fig:stati_bubbl} (a), the rectangular bubble deforms into a circle over time. In addition the figure indicates the temporary velocity vectors which decay in time. In the final state, the  pressure difference between the bubble centre and a far outside location reads $\Delta p = \SI{2.0}{Pa}$, which matches the result of the 2D Young-Laplace law, i.e. $\Delta p = \sigma_\mathrm{a,b} /R$, and also indicates the correct prediction of the final bubble radius $R = L / \sqrt{\pi}$. To underline the correct pressure approximation, the same situation is simulated with an increased transition length for three different surface tension values $\sigma_\mathrm{a,b} = 0.5, 0.25,0.0625$ [N/m] at $\mathrm{Ca}_\mathrm{L} = 0.1$, $\mathrm{Pe}_\mathrm{L} = 200 [N/m] / \sigma_\mathrm{a,b}$, $\mathrm{We} = 4 [N/m] / \sigma_\mathrm{a,b}$. Fig. \ref{fig:stati_bubbl} (b) shows the resulting pressure distributions over a radial coordinate which reveal a fair agreement with theoretical results.

%%%%%%%%%%%%%%%%%%%%%%%%%%%%%%%%%%%%%%%
\section{Mobility Parameter in Under-Resolved Flows}
\label{sec:appro_mobil_estim}
The section discusses means to model the mobility parameter in under-resolved flow simulations, where the surface tension influence is neglected due to the coarse resolution of the interface thickness $\gamma$. As outlined in Sect. \ref{sec:ch}, the neglect of surface tension yields $C_2=0$ and the concentration equation (\ref{equ:MIX2a}) simplifies towards
\begin{align}
\frac{\partial c}{\partial t} + \frac{\partial \, v_i c}{\partial x_i} = \frac{\partial}{\partial x_i} \left[ 
M \frac{\partial}{\partial x_\mathrm{i}} \left( C_1 \frac{\partial b}{\partial c} \right) \right] 
\; . 
\label{equ:simple}
\end{align}
Using $\nabla_i b= (\nabla_i c) \,  \partial b/\partial c$ together with the definition  of the double well potential $b$ (\ref{equ:CH-b}), the r.h.s. of (\ref{equ:simple}) reads  
\begin{align}
\frac{\partial}{\partial x_i} \left[ 
M \frac{\partial}{\partial x_\mathrm{i}} \left( C_1 \frac{\partial b}{\partial c} \right) \right] 
= \, 
 \frac{\partial}{\partial x_\mathrm{k}}\left[  M \, C_1 \left( 12c^2-12c+2 \right) \frac{\partial c}{\partial x_\mathrm{k}} \right]
= 
 \frac{\partial}{\partial x_\mathrm{k}}\left[  \nu_c \frac{\partial c}{\partial x_\mathrm{k}} \right]
\; . 
\label{equ:cahn_hilli_diffu}
\end{align}
Depending on the concentration value, (\ref{equ:cahn_hilli_diffu}) acts locally diffusive ($\nu_c \ge 0$) or compressive ($\nu_c <0$). The sign-change of the apparent viscosity $\nu_c=2 C_1 M(6c^2-6c+1)$ resembles compressive approximations of the convective term,  which switch from upwind to downwind approximations along the interface, and thus from positive to negative (apparent) viscosities, to keep the interface sharp  \cite{ubbink1999method, muzaferija1998computation}. The apparent viscosity in (\ref{equ:cahn_hilli_diffu}) zeroes at $c= 0.5 (1 \pm 1/\sqrt{3})$ and is negative over approximately 58\% of the transition region.
Aiming at a closure for the mobility parameter in under-resolved simulations, we separate the mobility into a physical and a modelled part, i.e. 
$
M= {M}^\text{phys} + {M}^\text{mod}.
$ 
The physical part is usually assigned to fairly small values, e.g $M^\text{phys} << \SI{1E-15}{m^3 s/kg}$ \cite{jacqmin2000contact,magaletti2013sharp}. Jacqmin \cite{jacqmin1999calculation} reports that the mobility typically scales with the transition length $M \propto {\gamma}^n$, where $n$ varies between $ 1 \leq n \leq 2$. Moreover, a recent publication of Magaletti et al. \cite{magaletti2013sharp} suggests $n=2$ and thus $Pe \sim Ca^{-1}$ or $\gamma \sim \sqrt{\sigma_{a,b}  M / V }$. Most engineering settings are therefore unable to sufficiently resolve the transition length and we consider the numerical contribution $M^\text{mod}$ to be dominant.

\subsection{Homogeneous  Mobility Model}
The  formulation of $M^\text{mod}$ is based on the interface blurring introduced by upwind-biased schemes. 
An estimation of the  tensorial numerical diffusion at a cell face returned by a first-order upwind scheme might read 
\begin{align}
\mathrm{\nu_{ij}^{\mathrm{UDS}} }^{(f)}
%(x_{k})  
=  (\lambda  \delta x_\mathrm{j} \, v_\mathrm{i} )^{(f)} 
%unsteady:   \left(1 + \delta t \left( \frac{v_\mathrm{i}}{\delta x_\mathrm{j}} \right)^{(f)} \right) = (\lambda  \delta x_\mathrm{j} \, v_\mathrm{i} )^{(f)} \left(1 + \mathrm{Co}^{(f)}_\mathrm{i,j} \right)
 \label{equ:lead_error} \, . 
\end{align}
Here $v_i^{(f)}$ denotes the velocity at the face center, $\delta x_j^{(f)}$ refers to the connecting vector from the upstream to the downstream adjacent cell center and
$\lambda \delta x_j^{(f)}$ approximates the distance between the face  and the upstream cell. Depending on the time discretization scheme, the related error might be included into the estimate of the mobility parameter from a modified equation analysis. For the example of a first-order implicit time discretization, the modified equation analysis suggests a simple supplement of a Courant number term,  i.e. $\mathrm{\nu_{ij}^{\mathrm{UDS}} }^{(f)}_{\rm unsteady}= \mathrm{\nu_{ij}^{\mathrm{UDS}} }^{(f)} (1+\mathrm{Co}_{\rm i,j}^{(f)})$.
The estimate (\ref{equ:lead_error})  is spatially and temporally variable. Spatially volatile mobility distributions are deemed to obstruct the robustness of the procedure.  Hence,  we confine our interest to homogenized approaches and estimate the  mobility based on the maximum norm of the matrix valued numerical diffusion
\begin{align}
 {M}^\text{mod}  =  \frac{\tilde{M}}{ C_1 \, \bigg| \left\lbrace n_\mathrm{f}: \delta x_\mathrm{i} \frac{\partial c}{\partial x_\mathrm{i}}^{(f)} \geq \delta_\mathrm{c} \right\rbrace \bigg| } \sum_{f}^{n_f} \begin{cases} \underset{\mathrm{i,j}}{ \mathrm{max}} \left(   |  
 \lambda  \delta x_\mathrm{j} \, v_\mathrm{i} | 
% unsteady:  (1 +\mathrm{Co}_\mathrm{i,j}
) 
 \right)^{(f)} & \text{if } \ \delta x_\mathrm{i} \frac{\partial c}{\partial x_\mathrm{i}}^{(f)} \geq \delta_\mathrm{c} \, ,  \\
0 & \text{otherwise \, .} \end{cases} \label{equ:diffu_error_estim}
\end{align}
Note that the field is filtered to extract the interface region, i.e. only faces with a  projected concentration gradient above $\delta_\mathrm{c} = 10^{-3}$ are considered. 
A von-Neumann stability analysis of the first-order discretized 1D equation at the  interface location ($c=0.5$) yields the following estimates for stable solutions (cf. appendix \ref{app:neumann}) 
\begin{align}
\tilde{M} \leq  \mathrm{1} \ \ \qquad  {\rm and } \qquad  \ \tilde{M} \geq  \left[1 + \frac{2}{\mathrm{Co} \left[ 1 - \mathrm{cos}(\varphi) \right]} \right], \label{equ:stable_mobility}
\end{align}
where $\varphi$ represents the phase angle. Approaching the steady state limit, the analysis only excludes $\tilde M=1$. In case of  $\varphi \to 0$, the branch $\tilde M \le 1$ might be a saver recommendation. Therefore $\tilde M=0.1$ is used for all  applications displayed in section \ref{sec:two_dimen}, for which  no stability problems are observed. If alternative convection schemes are used, the 1st-order upwind based analysis still provides reliable estimates. 

\begin{figure}[h]
\centering
\includegraphics[scale=1.3] {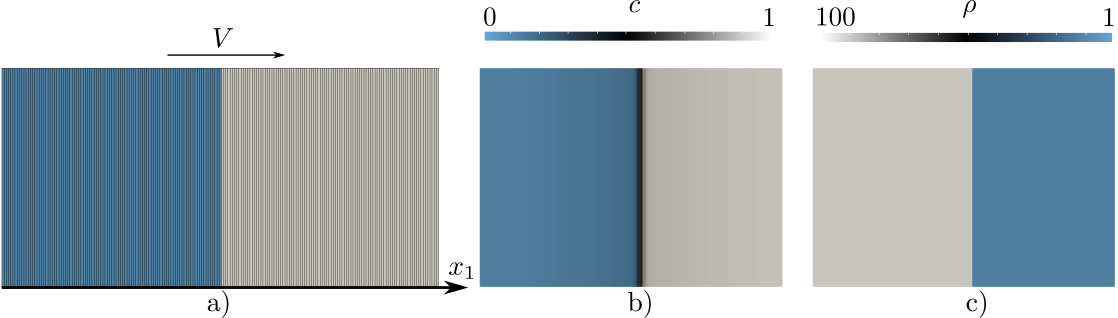}
\caption{(a) Computational grid with initial conditions for a 1D concentration advection, (b)  slightly blurred interface predicted by a CH-NS simulation using $\tilde{M} = 0.1$ and $\mathrm{Co} = 1$ after $t = \SI{5}{s}$ and (c)  corresponding density field 
returned by the non-linear equation of state (\ref{equ:m-func2}; $\gamma_\mathrm{m} = 0.05$).}
\label{fig:mater_inter}
\end{figure}

Further inspection  reveals, that under no circumstances the predicted phase transition spans less than five control volumes. 
Improved sharpening is obtained from the non-linear equation of state (\ref{equ:m-func2}). 
%
%
%%%%%%%%%%%%%%%%%%%%%%%%%%%%%%%%%%%%%%%%%%
An illustrative 1D example is used to demonstrate this. 
In this example, a free surface is transported by a prescribed flow field on the grid depicted in Fig. \ref{fig:mater_inter} (a). The horizontal flow field is directed from left to right with a  constant velocity of $v_\mathrm{1} = \SI{1}{m/s}$, and only the concentration equation (\ref{equ:CH1a}) is computed. The employed grid  is homogeneous ($\lambda = 0.5$) and features  $\delta x_1 = 10^{-3}$m and $\delta x_2=1$m.  The simulation is initialised with a sharp interface along a vertical line at the centre location $x_1 = 1$m.
Fig. \ref{fig:mater_inter} (b) displays a partly blurred interface from one CH simulation with $\tilde{M} = 0.1$ and $\mathrm{Co} = 1$ after $t = \SI{5}{s}$.
The corresponding density field obtained from the non-linear equation of state using $\gamma_\mathrm{m} = 0.05$ is displayed in Fig. \ref{fig:mater_inter} (c). 
Although the concentration field is slightly blurred, the resulting density and viscosity fields are sharp.

%%%%%%%%%%%%%%%%%%%%%%%%%%%%%%%%%%%%%%%%%%
A second 2D example refers to a circle of radius $r/L=0.15$ that is initially placed at $x_\mathrm{i}/L = [0.5,0.75]^\mathrm{T}$ in a square of unit length ($L=\SI{1}{m}$) as described in \cite{rider1998reconstructing}.
In this frequently employed validation example, the free surface is advected under a spatial deforming velocity field $v_\mathrm{1} = -\partial \psi / \partial x_\mathrm{2}$, $v_\mathrm{2} = \partial \psi / \partial x_\mathrm{1}$ where $\psi = \mathrm{sin}^2(x_\mathrm{1} \pi) \, \mathrm{sin}^2(x_\mathrm{2} \pi) / \pi$ defines the stream function.
% During the numerical process the velocity field is kept frozen.
%
Above all, we would like to show that the proposed approach allows for courant numbers $\mathcal{O}(1)$ and has resharpening capabilities also under complex and large interface deformations. Therefore, a comparatively fine numerical grid with $\delta x_\mathrm{i} / L = 1 / 400$ is used combined with 
$\delta t = \mathrm{Co} \,  \delta x_\mathrm{i} / V$ and $\mathrm{Co} = 2$ as well as $V = \SI{1}{m/s}$ in accordance to the maximum value from the stream function definition.
Symmetry boundary conditions are used and second order approximations are conducted for transient (ITTL), convective (QUICK) and diffusive (CD) terms.
Results are assessed by means of the predicted interface sharpness and the spatial progression of the interface over time. A local sharpness indicator quantity $q$ is used to judge the interface quality.
%, as suggested in \cite{manzke2018development}. 
The latter employs the computed concentration gradient at an interfacial face and multiplies this with twice the grid-spacing, i.e. $q^{-1}= 2 \delta x_\mathrm{i} \, (\partial c/\partial x_\mathrm{i})_{(c=0.5)}$. In the present example, an interfacial face is a face that is adjacent by one cell featuring $c<0.5$ and one cell featuring $c>0.5$.
% -- for more details we refer to \cite{manzke2018development}. 
A perfectly sharp interface results in $q = 1$, acceptable interfaces follow from $q < 3$ and a global reference is determined by the arithmetic average of all local sharpness indicators $q$, labeled $Q$. 

\begin{figure}[h]
\begin{minipage}{0.475\textwidth}
\centering
\input{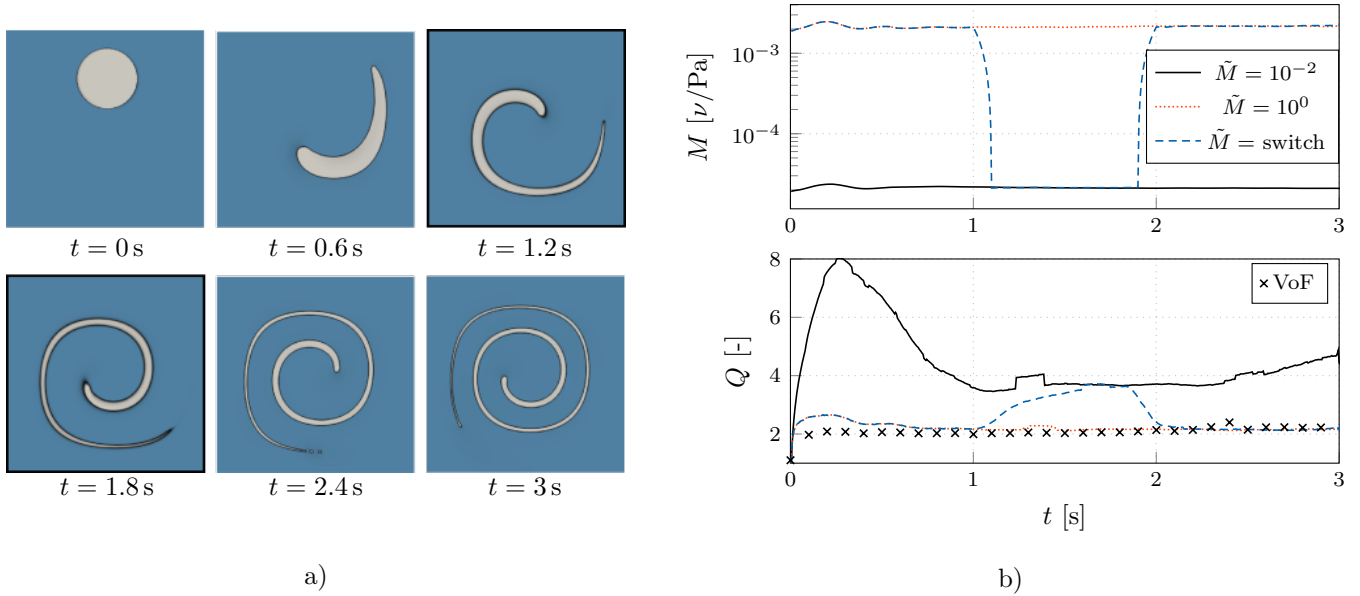}
\\
a)
\end{minipage}
\begin{minipage}{0.475\textwidth}
\raggedleft
\longBigPicture
\pgfplotsset{
	legend columns=1
}
\begin{tikzpicture}
\begin{axis}[
 ylabel style={text width=0.25\textwidth,align=center},
 ylabel={$M$ [$\nu / \mathrm{Pa}$]},
 xmin=0,xmax=3,
 ymode=log,
 legend style={at={(1.0,0.5)},anchor=east},
]
\addplot [line1] table[x expr={\thisrowno{0}},y expr={\thisrowno{1}}]{data/Kothe_Case_Mobility.dat};
\addplot [line2] table[x expr={\thisrowno{0}},y expr={\thisrowno{2}}]{data/Kothe_Case_Mobility.dat};
\addplot [line3] table[x expr={\thisrowno{0}},y expr={\thisrowno{3}}]{data/Kothe_Case_Mobility.dat};

\addlegendentry{$\tilde{M} = 10^{-2}$};
\addlegendentry{$\tilde{M} = 10^{0}$};
\addlegendentry{$\tilde{M} = $ switch};

\end{axis}
\end{tikzpicture}
\longBigPicture
\pgfplotsset{
	legend columns=2
}
\begin{tikzpicture}
\begin{axis}[
 ylabel style={text width=0.25\textwidth,align=center},
 xlabel={$t$ [s]},
 ylabel={$Q$ [-]},
 xmin=0,xmax=3,
 ymin=1.0,ymax=8.0,
 %ymode=log,
 %xtick={-10,12,34,56,78,100},
 %ytick={-0.01,-0.005,0,0.005,0.01},
 %yticklabels={-0.01,-0.005,0,0.005,0.01},
 %scaled y ticks = false
]
\addplot [mark1, only marks,mark repeat=200] table[x expr={\thisrowno{0}},y expr={\thisrowno{1}}]{data/Kothe_Case_Sharpness_VoF.dat};
\addplot [line1] table[x expr={\thisrowno{0}},y expr={\thisrowno{1}}]{data/Kothe_Case_Sharpness.dat};
\addplot [line2] table[x expr={\thisrowno{0}},y expr={\thisrowno{2}}]{data/Kothe_Case_Sharpness.dat};
\addplot [line3] table[x expr={\thisrowno{0}},y expr={\thisrowno{3}}]{data/Kothe_Case_Sharpness.dat};

\addlegendentry{VoF};

\end{axis}
\end{tikzpicture}
%\\
\centering
b)
\end{minipage}
\caption{Time evolution of (a) the concentration contour for a prescribed mobility of $\tilde{M} = 1$ that is reduced to
$\tilde{M} = 0.01$ between $\SI{1}{s} \leq t \leq \SI{2}{s}$
and (b) 
mobility $M$ (top) and global sharpness indicator $Q$ (bottom) over the simulation time for different under-resolved CH-NS
% and one VoF 
simulations at $\mathrm{Co} = 2$.}
%
%For the VoF case only a sharpening value is available.
\label{fig:kothe_case}
\end{figure}

Three CH-NS simualations are performed. Two simulations refer to temporally constant mobility parameter values $\tilde{M}=1$ and $\tilde{M}=0.01$, which should feature measurable differences on the predicted sharpness. In the third case, $\tilde{M}$  switches between the two constant values, i.e. $\tilde{M}(\SI{1}{s} \le t \le \SI{2}{s})=0.01$ and $\tilde{M}(t < \SI{1}{s}, \SI{2}{s} < t)=1$.  
%(Fig. \ref{fig:kothe_case} b) top).
%
The evolution of the advected concentration over time is displayed in Fig. \ref{fig:kothe_case} (a) for the switching mobility parameter case. In the first and last third of the simulation the interface remains practically sharp, which is no longer the case for the time $\SI{1}{s} \leq t \leq \SI{2}{s}$ where the lower choice of $\tilde{M}$ is not able to overcome the numerical diffusion. The visible temporary blurring is also displayed by the sharpness indicator in Fig. \ref{fig:kothe_case} (b) (bottom) that underlines the (on-the-fly) resharpening capability and the competent predictive performance of the CH-NS in comparison to VoF schemes (HRIC; $\mathrm{Co} = 0.2$). 

%%%%%%%%%%%%%%%%%%%%%%%%%%%%%%%%%%%%%%%%%%%
\section{Validations \& Applications}
\label{sec:two_dimen}

\subsection{Gravity and Capillary Wave}
\label{sec:gcwave}
\begin{figure}[h]
\centering
\subfigure[]{
\centering
\includegraphics[width=0.40\textwidth]{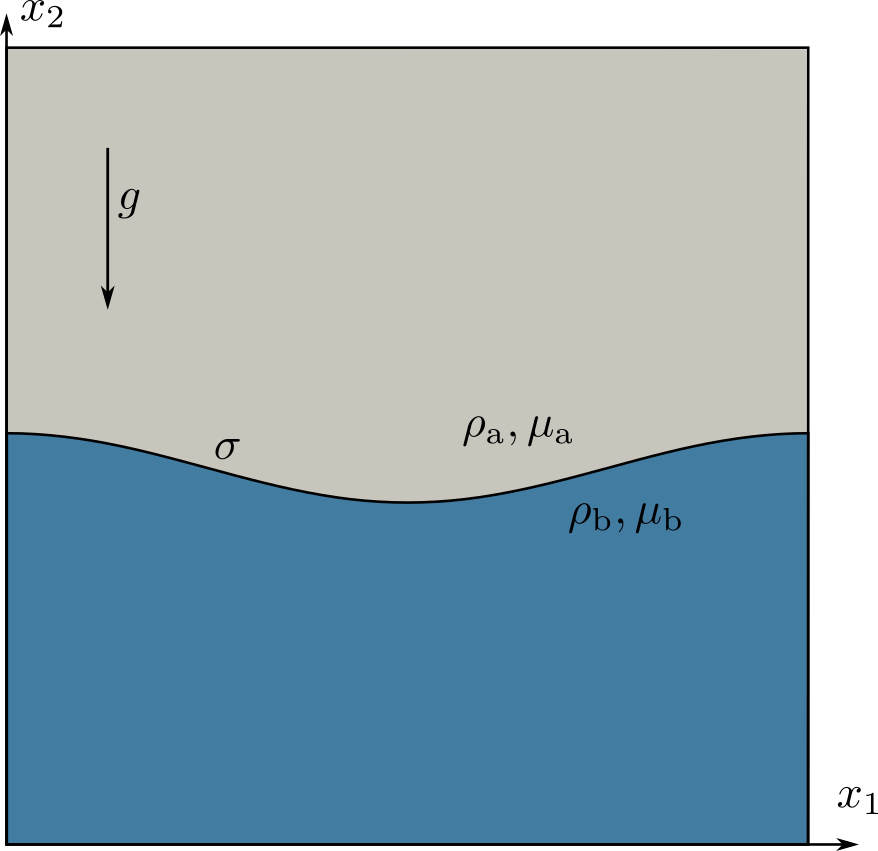}
}
\hspace{1cm}
\subfigure[]{
\centering
\includegraphics[width=0.375\textwidth]{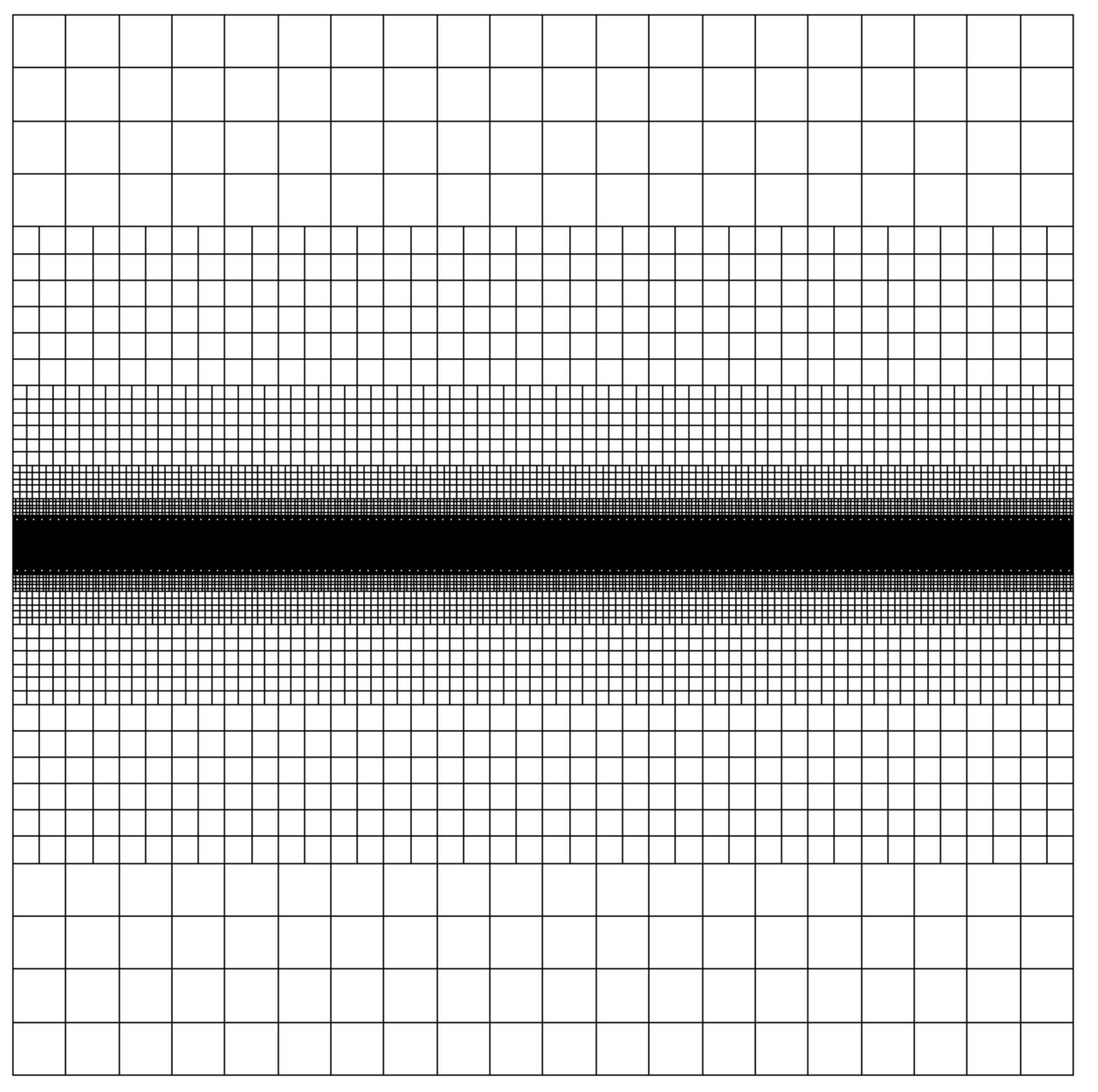}
}
\caption{(a) Initial setup of the standing wave case (scaled for visualization purposes) and (b) employed computational grid.}
\label{fig:standing_waves}
\end{figure}

The first example deals with the decay of standing waves which are initialised 
according to Fig.  \ref{fig:standing_waves}(a). We aim to assess, if the shear driven energy exchange between the two fluids, labelled a \& b,  is correctly captured in both, the capillary and the gravity regime under the influence of vertical gravitational acceleration $g_2$. Numerical results are compared with analytical solutions of Prosperetti (1981), which exist for identical kinematic viscosities ($\mu_\mathrm{a} / \rho_\mathrm{a} = \mu_\mathrm{b} / \rho_\mathrm{b}$) in the linear (laminar) flow regime.
The initial wave length complies with a unit wave number ($k = 2 \pi / \lambda = 1$) and the initial wave amplitude corresponds to $a =0.01  \lambda$. The reference velocities refer to $V = \sqrt{|g_\mathrm{2}| \, \lambda}$ and 
%FALSCH: $V = \sigma_\mathrm{a,b} / \rho_\mathrm{b}$  
$V =\sqrt{ \sigma_\mathrm{a,b} / (\lambda \, \rho_\mathrm{b}} )$
for the gravity and the capillary case, respectively. 
The extent of the 2D computational domain depicted in Fig. \ref{fig:standing_waves} (b) reads $ \lambda \times \lambda$. A locally refined grid with approximately 250.000 isotropic control volumes is employed. The resolution of the  free surface region refers to $\delta x_\mathrm{1} = \delta x_\mathrm{2} = \lambda / 4000$. The time step is assigned to $\delta t= \lambda /(4000 V)$ which is sufficient to ensure Courant numbers below $\mathrm{Co} < 0.1$.  Symmetry (no-slip) conditions are used along constant $x_\mathrm{1}$ ($x_\mathrm{2}$) boundaries of the domain. 
%VoF results refer to a compressive HRIC scheme of \cite{muzaferija1998computation}.

\begin{figure}[htbp]
\centering
\longPicture
\pgfplotsset{
	legend columns=2
}
\begin{tikzpicture}
\begin{axis}[
 ylabel style={text width=0.25\textwidth,align=center},
 xlabel={$t/ \sqrt{\lambda / g}$ [-]},
 ylabel={$x_\mathrm{2,FS}/\lambda$ [-]},
 xmin=0,xmax=15,
 ymin=-0.01,ymax=0.01,
 ytick={-0.01,-0.005,0,0.005,0.01},
 yticklabels={-0.01,-0.005,0,0.005,0.01},
 scaled y ticks = false
]

\addplot [line1] table[x expr={\thisrowno{0}},y expr={\thisrowno{1}}]{data/Gravity_Wave_FreSCo_HRIC_Results.dat};
\addplot [line3] table[x expr={\thisrowno{0}},y expr={\thisrowno{1}}]{data/Gravity_Wave_FreSCo_mp00_Results.dat};
\addplot [mark1, only marks,mark repeat=10] table[x expr={\thisrowno{0}},y expr={\thisrowno{1}}]{data/Gravity_Wave_Prosperetti_Results.dat};
%\addplot [mark1, only marks,mark repeat=1] table[x expr={\thisrowno{0}},y expr={\thisrowno{1}}]{data/Gravity_Wave_KTH_Results.dat};

%\addplot [line4] table[x expr={\thisrowno{0}},y expr={\thisrowno{1}}]{data/Gravity_Wave_FreSCo_mm02_Results.dat};
%\addplot [line5] table[x expr={\thisrowno{0}},y expr={\thisrowno{1}}]{data/Gravity_Wave_FreSCo_mp02_Results.dat};

\addlegendentry{VoF};
\addlegendentry{CH-NS};
%\addlegendentry{Prosperetti (1981)};
%\addlegendentry{Akervik \& Vartdal (2016)};

%\addlegendentry{CH-NS ($\tilde{M} = 10^{-2}$)};
%\addlegendentry{CH-NS ($\tilde{M} = 10^{+2}$)}; 

\end{axis}
\end{tikzpicture}
\vspace{-5mm}
\longPicture
\pgfplotsset{
	legend columns=2
}
\begin{tikzpicture}
\begin{axis}[
 ylabel style={text width=0.25\textwidth,align=center},
 xlabel={$t/ \sqrt{\rho_\mathrm{b} \lambda^3/\sigma}$ [-]},
 ylabel={$x_\mathrm{2,FS}/\lambda$ [-]},
 xmin=0,xmax=3,
 ymin=-0.01,ymax=0.01,
 ytick={-0.01,-0.005,0,0.005,0.01},
 yticklabels={-0.01,-0.005,0,0.005,0.01},
 scaled y ticks = false
]
\addplot [mark1, only marks,mark repeat=10] table[x expr={\thisrowno{0}},y expr={\thisrowno{1}}]{data/Cappilary_Wave_Prosperetti_Results.dat};
\addplot [line3,each nth point={10}] table[x expr={\thisrowno{0}},y expr={\thisrowno{1}}]{data/Cappilary_Wave_FreSCo_Results.dat};
%\addplot [mark1, only marks,mark repeat=1] table[x expr={\thisrowno{0}},y expr={\thisrowno{1}}]{data/Cappilary_Wave_Chen_Results.dat};

%\addlegendentry{CH-NS (Lin.)};
\addlegendentry{Prosperetti (1981)};
%\addlegendentry{Chen et al. (1999)}; 

\end{axis}
\end{tikzpicture}
\vspace{5mm}
\caption{Comparison of the  analytical (symbols) and numerical time evolution for the wave elevation  at the left boundary ($x_1=0$) obtained for the gravity (top) and the capillary (bottom) case.}
\label{fig:standing_waves_results}
\end{figure}
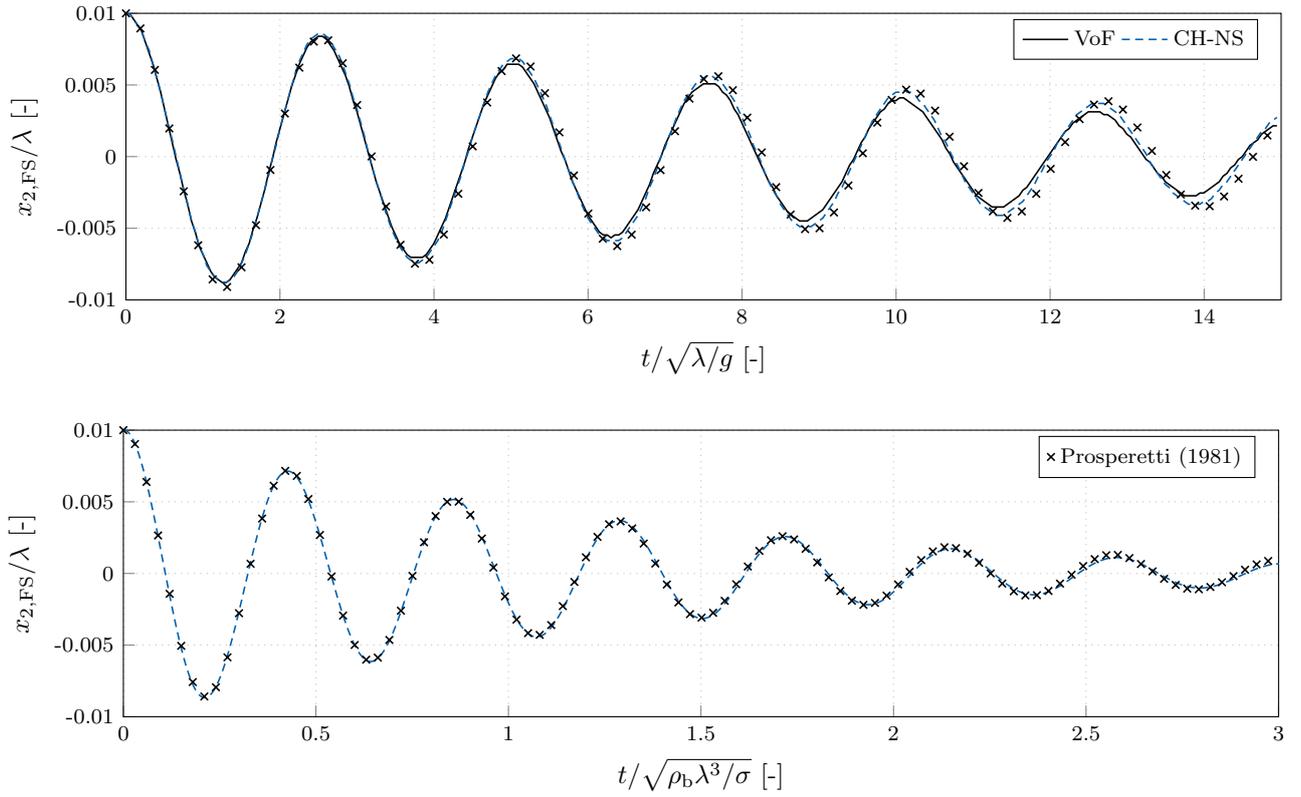

For the gravity wave, surface tension influences are neglected and the density ratio and Reynolds-number read $\rho_\mathrm{a} / \rho_\mathrm{b} = 1/100$ and $\mathrm{Re} = V \lambda / \nu_\mathrm{b} = 1000$. The CH-NS simulations are based on the mobility estimation described in section \ref{sec:appro_mobil_estim} and the  non-linear material law ($\tilde{M}=0.1$, $\gamma_\mathrm{m} = 0.1$). Figure \ref{fig:standing_waves_results} (top) displays the evolution of the free-surface elevation at the horizontal left end ($x_\mathrm{1} = 0$) predicted by CH-NS and VoF next to the analytical solution of Prosperetti.
%as well as numerical PLIC-VoF results of Akervik and Vartdal \cite{akervik2016simulation}. 
In comparison to the linear theory, the top figure reveals a slightly stronger wave damping and a minor phase shift returned by the CH-NS approach for the gravity case, which outperforms the present VoF method.
The latter is a consequence of the resharpening character of the CH-NS approach.

For the capillary case, the density and viscosity ratios read $\rho_\mathrm{a} / \rho_\mathrm{b} = 1 / 100$ and $\mu_\mathrm{a} / \mu_\mathrm{b} = 1 / 10$ respectively, the interface thickness is resolved by 10 vertical cells and follows from a Cahn-number of $\mathrm{Ca}_\mathrm{\lambda} = \gamma/\lambda = 1 / 400$. The Ohnesorge-number adjusts the surface tension force $\sigma$ and reads 
$\mathrm{Oh} = \mu_\mathrm{b} / \sqrt{ (\sigma_\mathrm{a,b} \lambda \rho_\mathrm{b})} = 1 / 100$. The mobility parameter is prescribed in accordance with a Peclet-number of $\mathrm{Pe}_\mathrm{\lambda} = 2 \cdot 10^{10}$ together with the linear material law. Conclusions drawn for the capillary case are similar to the gravity case, as indicated by Figure \ref{fig:standing_waves_results} (bottom), which compares the evolution of the wave amplitude at the horizontal left end ($x_\mathrm{1} / \lambda = 0$) predicted by the present CH-NS  with the analytical solution of Prosperetti .
%as well as numerical (PLIC-VoF) results of Chen et al. \cite{chen1999two}.

\subsection{Submerged Hydrofoil}
\label{sec:foil}
\begin{figure}[h]
\centering
\subfigure[]{
\centering
\includegraphics[scale=0.265]{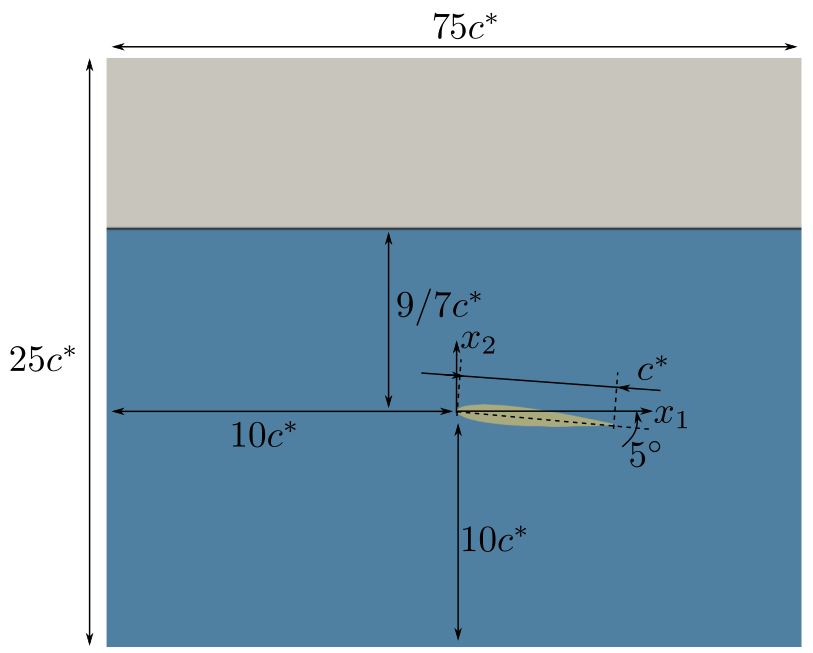}
}
\hspace{1cm}
\subfigure[]{
\centering
\includegraphics[scale=1]{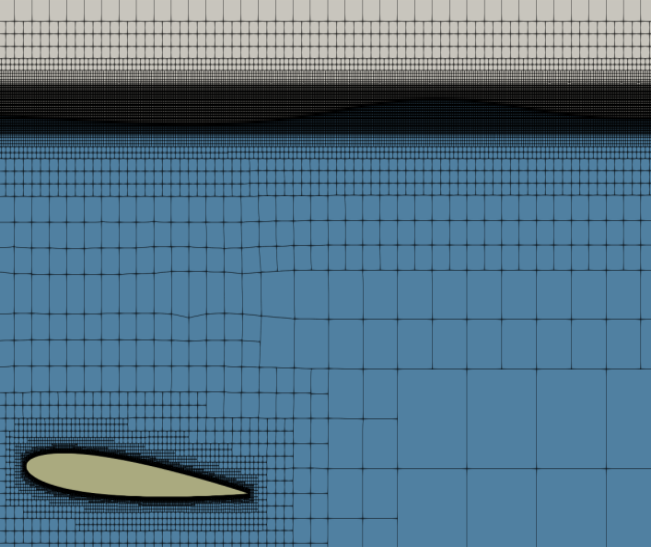}
}
\caption{(a) Schematic drawing of the initial configuration and (b) unstructured numerical grid around the foil and the free surface.}
\label{fig:dunca_foil}
\end{figure}

The second example refers to the wave pattern downstream of a submerged NACA0012 hydrofoil at  $\SI{5}{\degree}$ incidence in accord with experimental data of Duncan \cite{duncan1981experimental,duncan1983breaking}, cf. Fig. \ref{fig:dunca_foil}(a). The chord length to submergence ratio at the leading edge of the foil reads $c/L= 7/9$. The study is performed for a turbulent flow at $\mathrm{Re} = v_1 c/\nu_b = \SI{144 855}{}$ and $\mathrm{Fn} = v_1/\sqrt{GL} = \SI{0.567}{}$, based on the gravitational acceleration $G$, the inflow velocity $v_1$ and the kinematic viscosity of the water $\nu_b$. The two-dimensional domain has a length and height of $75 c$ and $25 c$, where the inlet and bottom boundaries are located 10 chord-lengths away from the origin. A dimensionless wave length of $\lambda^* = \lambda/ L = 2 \, \pi \, \mathrm{Fn}^2 = 2.0193$ is expected. 

The utilized unstructured numerical grid is displayed in Fig. \ref{fig:dunca_foil} (b) and consists of approximately $\SI{150000}{}$ control volumes. The fully turbulent simulations employ a wall-function based $k-\omega$ SST model \cite{menter2003ten} and all convective terms are approximated using the QUICK scheme. At the inlet, velocity and concentration values are prescribed, slip walls are used along the top and bottom boundaries and a hydrostatic pressure boundary is employed along the outlet. The wall normal distance of the first grid layer reads $y^+ \approx \SI{30.0}{}$ and the free surface refinement employs approximately  $\delta x_\mathrm{1} / \lambda = 1/100$ cells in the longitudinal as well as $\delta x_\mathrm{2} / \lambda = 1/400$ cells in the normal direction. The VoF approach is integrated in pseudo time with a time step size of  $\delta t =\delta x_\mathrm{2} / V \ \mathrm{Co}$ together with $\mathrm{Co} = 0.1$.
% and refer to a compressive HRIC scheme of \cite{muzaferija1998computation}. 
The CH-NS results are obtained from a steady state approach.

\begin{figure}[h]
\centering
\subfigure[]{
\centering
\smallPicture
\pgfplotsset{
	legend columns=2
}
\begin{tikzpicture}
\begin{axis}[
 ylabel style={text width=0.25\textwidth,align=center},
 xlabel={$ x_\mathrm{1} / c $ [-]},
 ylabel={$ x_\mathrm{2,FS} / c $ [-]},
 xmin=-0.0,xmax=7.0,
 ymin=-0.06,ymax=0.06,
 legend style={at={(0.02,0.98)},anchor=north west},
 ytick={-0.06,-0.03,0,0.03,0.06},
 yticklabels={-0.06,-0.03,0,0.03,0.06},
 scaled y ticks = false
]

\addplot [line1] table[x expr={\thisrowno{0}},y expr={\thisrowno{1}}]{data/Duncan_Foil_Without_TL_HRIC.dat};
\addplot [line3] table[x expr={\thisrowno{0}},y expr={\thisrowno{1}}]{data/Duncan_Foil_Without_TL_CH_M_1E-01_nonlinear.dat};
\addplot [line2] table[x expr={\thisrowno{0}},y expr={\thisrowno{1}}]{data/Duncan_Foil_Without_TL_CH_M_1E-01_linear.dat};
\addplot [mark1,mark repeat=1,only marks] table[x expr={\thisrowno{0}},y expr={\thisrowno{1}}]{data/Duncan_Foil_Experimental_Data.dat};

\addlegendentry{VoF (Lin.)};
\addlegendentry{CH-NS (N.-Lin.)}; 
\addlegendentry{CH-NS (Lin.)};
\addlegendentry{Duncan (1981)};
 
\end{axis}
\end{tikzpicture}
}
\hspace{1mm}
\subfigure[]{
\centering
\smallPicture
\pgfplotsset{
	legend columns=3
}
\begin{tikzpicture}
\begin{axis}[
 ylabel style={text width=0.25\textwidth,align=center},
 xlabel={$ t_\mathrm{sim.} $ [h]},
 ylabel={$ \frac{2 |F_\mathrm{1}|}{\rho_\mathrm{b} V^1 c^2} $ [-]},
 %xmin=1E-04,xmax=1E+02,
 %ymin=1E+01,ymax=1E+02,
 xmode=log,
 ymode=log,
 legend style={at={(0.98,0.98)},anchor=north east},
]

\addplot [line1,each nth point=10] table[x expr={\thisrowno{0}},y expr={\thisrowno{1}}]{data/Duncan_Foil_Without_TL_forceX_HRIC.dat};
\addplot [line3,each nth point=10] table[x expr={\thisrowno{0}},y expr={\thisrowno{1}}]{data/Duncan_Foil_Without_TL_forceX_CH_Linear.dat}; 

\addlegendentry{VoF (Lin.)};
\addlegendentry{CH-NS (N.-Lin.)}; 
  
\end{axis}
\end{tikzpicture} 
}
\caption{Submerged hydrofoil case (Fn=0.567); comparison of predicted normalised (a) wave elevation and (b) drag force over wall-clock time.}
\label{fig:dunca_resul}
\end{figure}
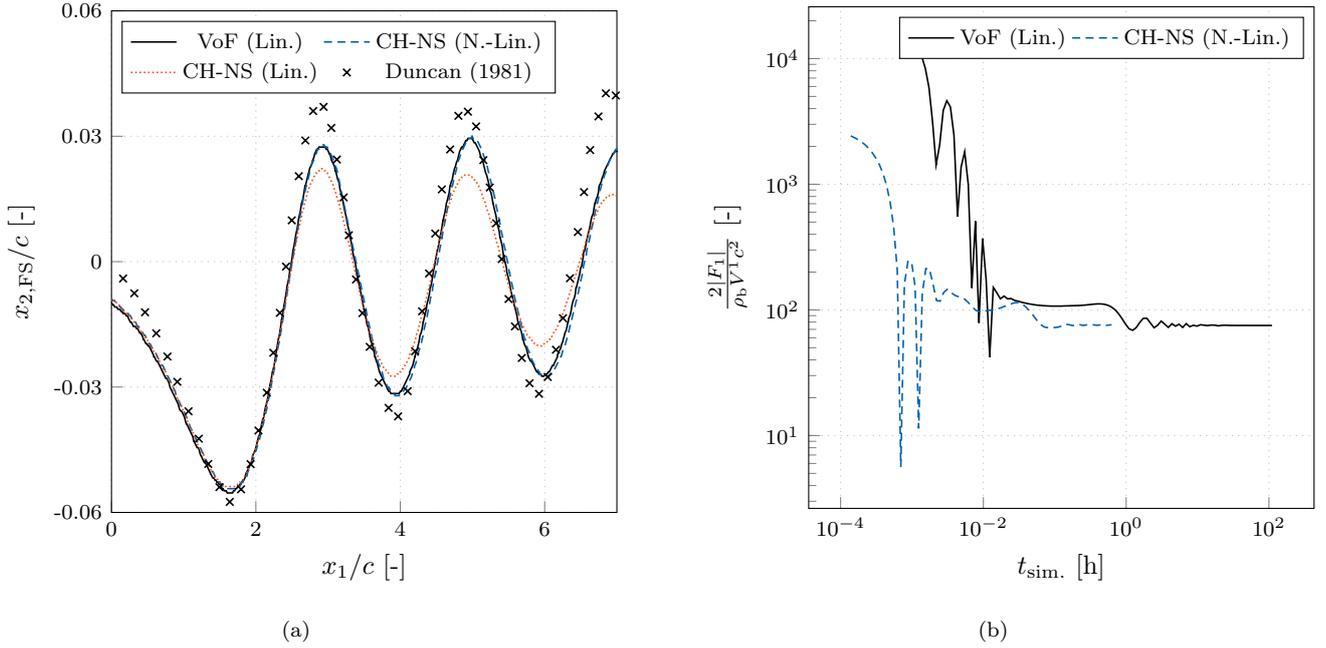

The  study neglects surface tension due to an under-resolved interface thickness and employs both the linear (\ref{equ:m-func1}) as well as the non-linear EoS (\ref{equ:m-func2}; $\gamma_\mathrm{m} = 0.1$). The surface tension force is neglected in the momentum equation and the concentration equation utilizes $C_\mathrm{2}=\SI{0}{N}$ and $C_\mathrm{1}=\SI{1}{Pa}$.
Fig. \ref{fig:dunca_resul}(a) shows the wave elevation for two CH-NS simulations with the same modelled mobility parameter ($\tilde{M} = 0.1$) but different EoS next to  the result of a VoF simulation. 
The non-linear EoS outperforms the linear version and drives the CH-NS approach closer to the experimental data as well as to the VoF result. Similar to the results displayed in Fig. \ref{fig:mater_inter}, the linear model provides slightly blurred density fields which translates into a reduction of the wave amplitude. 
Fig. \ref{fig:dunca_resul} (b) tracks the drag force coefficient over the simulation time $t_\mathrm{sim}$ for the VoF and the non-linear CH-NS simulations. The predicted drag differs about 0.4\% and a speed up of approximately one order of magnitude is achieved through the Courant-number independent CH-NS approach.

\subsection{Flow around a KCS container vessel}
\label{sec:three_dimen}
The final application  refers to the fully turbulent flow around an unappended Kriso container ship hull (KCS). Experimental resistance data and wave fields are published by \cite{kim2001measurement}  for a 1:31.6 scale  model and a large amount of comparative numerical data exists, e.g. \cite{larsson2010, kroger2018adjoint, kroger2016numerical,  manzke2013sub, banks2010free}. The distance between aft  and front perpendiculars of the hull model serves as a reference length $L = \SI{7.2786}{m}$ ($= L_\mathrm{pp}$). Other reference values refer to the gravity acceleration  $G$,  the inflow velocity magnitude $V$ and the kinematic viscosity of the water $\nu_b$. The model scale investigations refer to Reynolds- and Froude-numbers of $\mathrm{Re} = VL/\nu_b = \SI{1.4E+07}{}$ and $\mathrm{Fn}= V/\sqrt{GL} = \SI{0.26}{}$. The hull is fixed at the full scale static draught with zero trim and the motion and propulsion of the ship are suppressed during the simulation and the experiments. 

\begin{figure}[h]
\centering
\includegraphics[scale=0.8]{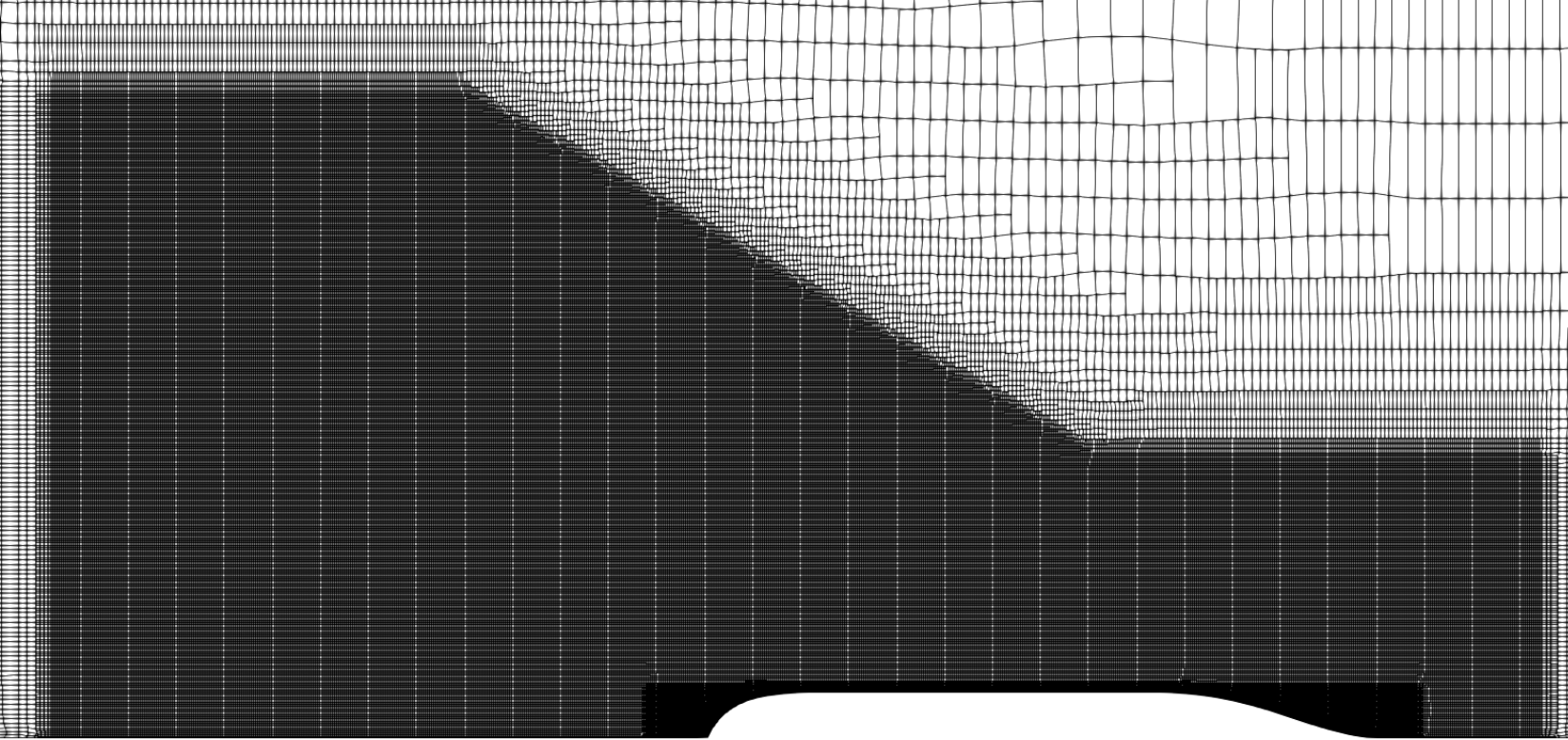}
\caption{Illustration of the employed computational mesh along the still water plane in the vicinity of the container ship hull.}
\label{fig:KCS_Mesh}
\end{figure}

The numerical grid consists of approximately 14.6 million unstructured hexahedral cells. The domains extends over 5L, 1.75L, 2.5L in longitudinal ($x_\mathrm{1}$), lateral ($x_\mathrm{2}$) and vertical ($x_\mathrm{3}$) direction. Due to symmetry only half of the flow field is simulated. The inlet is located upstream at $x_\mathrm{1} / L_\mathrm{pp} = 3$ and the free surface is initialized at $x_\mathrm{3} / L_\mathrm{pp} = 1.75$ over the lower boundary of the domain.
The surface of the hull is discretized with approximately 300,000 surface elements. The wall normal resolution of the hull refers to a dimensionless wall distances of $30 \leq y^+ \leq 100$ and justifies the use of wall functions. The vertical resolution of the free surface region is constant throughout the domain and attempts to resolve the expected wave amplitude of $ \SI{5E-04}{} L_\mathrm{pp} $ by hundred cells in the immediate vicinity of the hull. The tangential resolution of the free surface is refined within a Kelvin-Wedge to capture the resulting wave pattern. Based on the current Froude-number a dimensionless wavelength of $\lambda / L_\mathrm{pp} =  2 \ \pi \ \mathrm{Fn}^2 = 0.4247$  is  expected, which is approximated with roughly 100 cells.
Fig. \ref{fig:KCS_Mesh} indicates the different refinement levels for the near and the far field.

At the inlet, outlet, outer and lower boundaries, Dirichlet values for velocity and concentration are specified, while the pressure is extrapolated.  A reverse situation is given at the top face which corresponds to a pressure boundary. Symmetry and wall boundary conditions are declared along the midship plane as well a the hull. Turbulence is modelled by a high-Re $k - \epsilon$ model  \citep{wilcox1998turbulence}. Convective momentum transport is realised by a monotonicity-preserving QUICK scheme.
Similar to the hydrofoil case, data obtained from CH-NS simulation is compared with VoF results. CH-NS calculations refer again to steady simulations using $\tilde{M} = 0.1$ and the non-linear EoS. VoF calculations employ time stepping based on $\delta t = \delta x_\mathrm{3,FS} / V \ \mathrm{Co}_\mathrm{\delta x}$, where the Courant number is assigned to $\mathrm{Co} = 0.3$ and $\delta x_\mathrm{3,FS}$ denotes the vertical resolution of the free surface.
All simulations are performed until the integrated forces on the hull converge. 
% under the same HPC-conditions (e.g. amount and type of CPU).

\begin{figure}[h]
\centering
\smallPicture
\pgfplotsset{
	legend columns=3
}
\begin{tikzpicture}
\begin{axis}[
 ylabel style={text width=0.25\textwidth,align=center},
 xlabel={$ t_\mathrm{sim.} $ [h]},
 ylabel={$ \frac{2 |F_\mathrm{1}|}{\rho_\mathrm{b} V^2 A_\mathrm{wetted}} $ [-]},
 %xmin=1E-02,xmax=1E+02,
 %ymin=1E+01,ymax=1E+02,
 xmode=log,
 ymode=log,
 legend style={at={(0.98,0.98)},anchor=north east},
]

\addplot [line1] table[x expr={\thisrowno{0}},y expr={\thisrowno{1}}]{data/KCS_forceX_HRIC.dat};
\addplot [line3] table[x expr={\thisrowno{0}},y expr={\thisrowno{1}}]{data/KCS_forceX_CH_Linear.dat};

\addlegendentry{VoF (Lin.)};
\addlegendentry{CH-NS (N.-Lin.)};
 
\end{axis}
\end{tikzpicture} 
\caption{Evolution of the predicted drag force over the wall-clock simulation time for the VoF and CH-NS simulation of the container vessel at Re=1.4$\times$10$^{7}$ and Fn=0.26.}
\label{fig:KCS_force_data}
\end{figure}
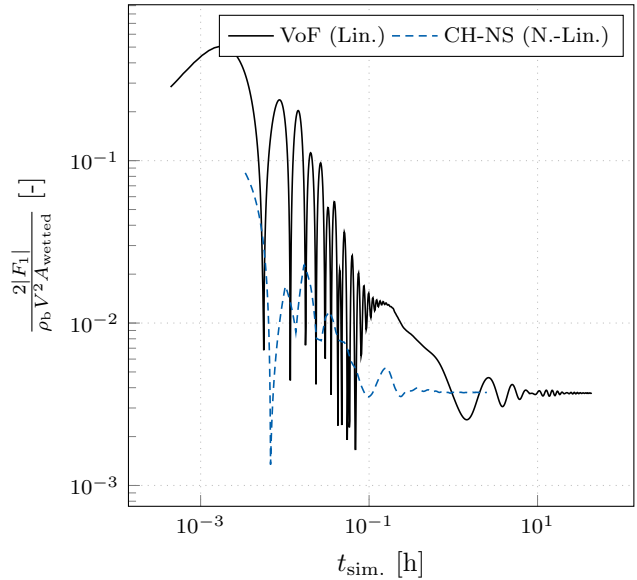

\begin{figure}[h]
\centering
\includegraphics[scale=1.0]{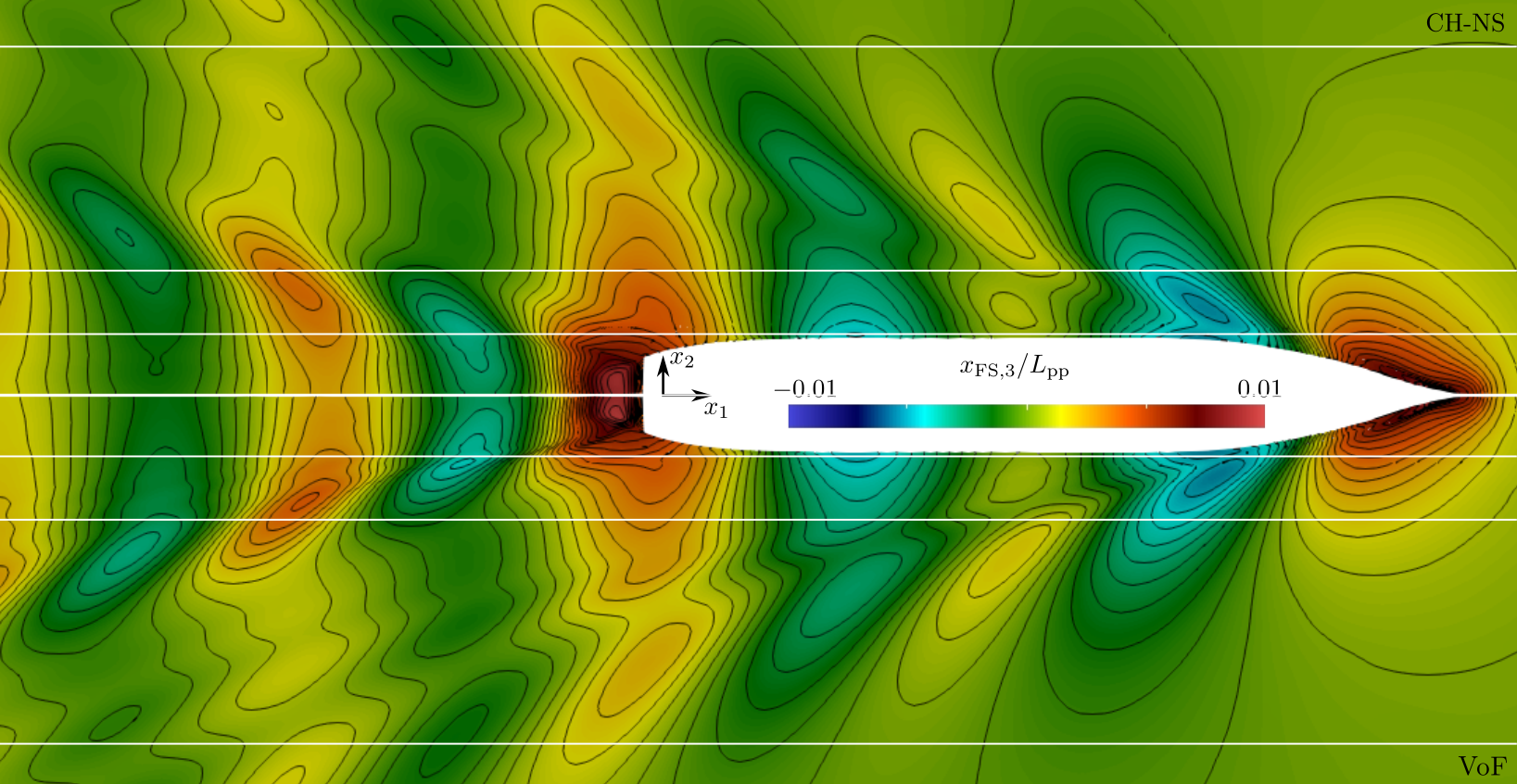}
\caption{Comparison of predicted wave field around the Kriso container vessel at Re=1.4$\times$10$^{7}$ and Fn=0.26 obtained by the VoF (bottom) and CH-NS (top) approach. White horizontal lines indicate evaluation planes used for the wave cuts displayed in Fig. \ref{fig:KCS_FS_slices}
.}
\label{fig:KCS_wave_field}
\end{figure}

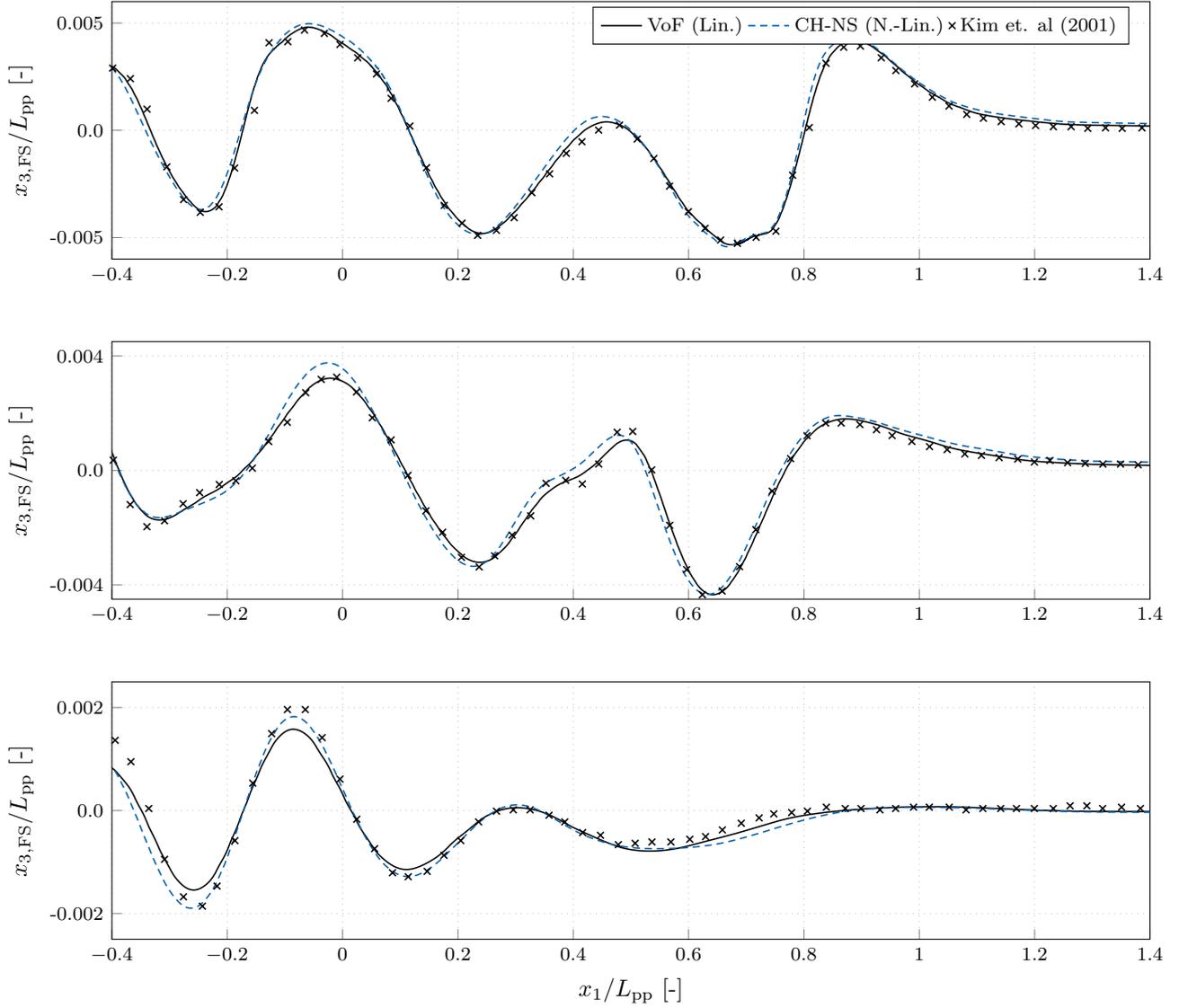
\begin{figure}[h]
\centering
\longPicture
\pgfplotsset{
	legend columns=3
}
\begin{tikzpicture}
\begin{axis}[
 ylabel style={text width=0.25\textwidth,align=center},
 ylabel={$ x_\mathrm{3,FS} / L_\mathrm{pp} $ [-]},
 xmin=-0.4,xmax=1.4,
 ymin=-0.006,ymax=0.006,
 legend style={at={(0.98,0.98)},anchor=north east},
 ytick={-0.005, 0.0, 0.005},
 yticklabels={-0.005, 0.0, 0.005},
 scaled y ticks = false
]

\addplot [line1] table[x expr={\thisrowno{0}},y expr={\thisrowno{1}}]{data/KCS_waterline_1_HRIC.dat};
\addplot [line3] table[x expr={\thisrowno{0}},y expr={\thisrowno{1}}]{data/KCS_waterline_1_CHNS.dat};
\addplot [mark1,mark repeat=1,only marks] table[x expr={\thisrowno{0}},y expr={\thisrowno{1}}]{data/KCS_waterline_1_EXP.dat};

\addlegendentry{VoF (Lin.)};
\addlegendentry{CH-NS (N.-Lin.)}; 
\addlegendentry{Kim et. al (2001)};
 
\end{axis}
\end{tikzpicture}

%%%%%%%%%%%%%%%%%%%%%%%%%%%

\pgfplotsset{
	legend columns=3
}
\begin{tikzpicture}
\begin{axis}[
 ylabel style={text width=0.25\textwidth,align=center},
 ylabel={$ x_\mathrm{3,FS} / L_\mathrm{pp} $ [-]},
 xmin=-0.4,xmax=1.4,
 ymin=-0.0045,ymax=0.0045,
 ytick={-0.004, 0.0, 0.004},
 yticklabels={-0.004, 0.0, 0.004},
 scaled y ticks = false
]

\addplot [line1] table[x expr={\thisrowno{0}},y expr={\thisrowno{1}}]{data/KCS_waterline_2_HRIC.dat};
\addplot [line3] table[x expr={\thisrowno{0}},y expr={\thisrowno{1}}]{data/KCS_waterline_2_CHNS.dat};
\addplot [mark1,mark repeat=1,only marks] table[x expr={\thisrowno{0}},y expr={\thisrowno{1}}]{data/KCS_waterline_2_EXP.dat};

\end{axis}
\end{tikzpicture}

%%%%%%%%%%%%%%%%%%%%%%%%%%%

\pgfplotsset{
	legend columns=3
}
\begin{tikzpicture}
\begin{axis}[
 ylabel style={text width=0.25\textwidth,align=center},
 xlabel={$ x_\mathrm{1} / L_\mathrm{pp} $ [-]},
 ylabel={$ x_\mathrm{3,FS} / L_\mathrm{pp} $ [-]},
 xmin=-0.4,xmax=1.4,
 ymin=-0.0025,ymax=0.0025,
 legend style={at={(0.98,0.98)},anchor=north east},
 ytick={-0.002, 0.0, 0.002},
 yticklabels={-0.002, 0.0, 0.002},
 scaled y ticks = false
]

\addplot [line1] table[x expr={\thisrowno{0}},y expr={\thisrowno{1}}]{data/KCS_waterline_3_HRIC.dat};
\addplot [line3] table[x expr={\thisrowno{0}},y expr={\thisrowno{1}}]{data/KCS_waterline_3_CHNS.dat};
\addplot [mark1,mark repeat=1,only marks] table[x expr={\thisrowno{0}},y expr={\thisrowno{1}}]{data/KCS_waterline_3_EXP.dat};

\end{axis}
\end{tikzpicture}
\caption{Comparison of measured and predicted wave elevation in 3 lateral planes, i.e. close to the hull at (a) $x_\mathrm{2} / L_\mathrm{pp} = 0.0741$, (b) at $x_\mathrm{2} / L_\mathrm{pp} = 0.1509$ and at a remote  position (c) $x_\mathrm{2} / L_\mathrm{pp} = 0.4224$.}
\label{fig:KCS_FS_slices}
\end{figure}

\begin{figure}[h]
\centering
\includegraphics[scale=1.0]{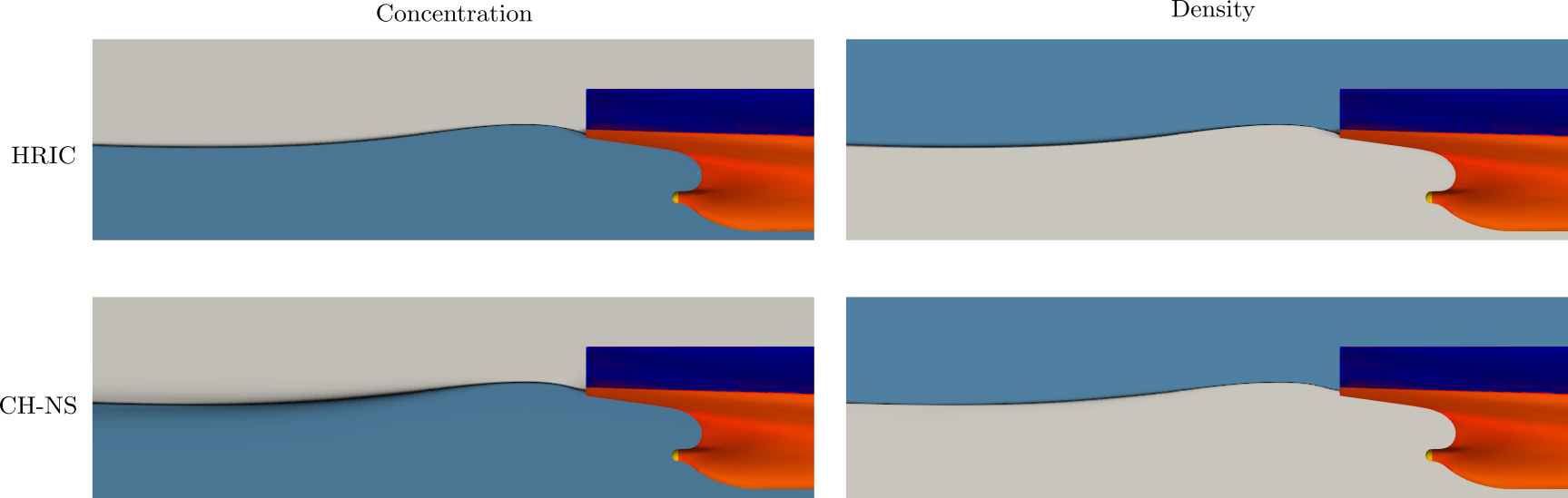}
\caption{Concentration (left) and  density field (right) obtained from a VoF  with linear EoS (top) and a CH-NS   with non-linear EoS (bottom).}
\label{fig:KCS_vs_CH}
\end{figure}

Fig. \ref{fig:KCS_force_data} depicts the evolution of total resistance over the wall-clock time. The predicted total drag force coefficient is normalised with the  static wetted surface of 9.5121 m$^2$ and converges to  $C_T=3.68 \times 10^{-3}$ and $C_T=3.66 \times 10^{-3}$ for the CH-NS and the VoF simulation.  Both values differ by only 0.5\% and compare favourable with 
the experimental value $C_T=3.56 \times 10^{-3}$ -- subject to the influence of other aspects, e.g. turbulence modelling. However, the CH-NS approach clearly outperforms the VoF simulation with respect to computational time, while introducing only minor additional wave damping, cf. Fig. \ref{fig:KCS_wave_field}. 
The wave elevation ($x_\mathrm{FS,3} / L_\mathrm{pp} $) measured at three different lateral planes through the free surface, i.e. $x_\mathrm{2} / L_\mathrm{pp} = 0.0741$, $x_\mathrm{2} / L_\mathrm{pp} = 0.1509$ and $x_\mathrm{2} / L_\mathrm{pp} = 00.4224$, is compared with experimental data in Fig. \ref{fig:KCS_FS_slices}. The predictive discrepancy is generally small and the non-linear CH-NS tends to provide slightly larger amplitudes.  Mind that the  non-linear EoS  \ref{equ:m-func2} leads to a significant sharpening of the density field, as illustrated by Fig. \ref{fig:KCS_vs_CH}. 

\section{Discussion and Outlook}
\label{sec:outlo}
The paper presents an alternative approach for the simulation of marine free surface flows at engineering scale. The method (labelled CH-NS) can be displayed as a Volume-of-Fluid (VoF) approach which is extended by a diffusive right hand side of order four obtained from a Cahn-Hilliard (CH) system. While the CH-framework is often used to describe phase separation processes along diffuse interfaces, the phase transition region usually falls below the typical grid resolution in  engineering settings. Therefore these settings closely resemble the sharp interface limit.

The phase separating characteristics of the CH-NS system is supported by negative diffusivity of the concentration equation in the central transition regime. This involves $\approx 60 \%$ negative diffusion for the double-well potential used in this paper, and scales with the mobility. A spatially homogeneous but temporally/iteratively variable mobility, adapted to an error expression, is suggested. This leads to robust results -- based on traditional upwind biased convective approximations --  with a fair predictive accuracy and resharpening capabilites. The model involves a free parameter which is assigned to a value well below the stability limit. An alternative non-linear material law is supplementary used to improve minimal blurring. 

The implementation of the CH-NS system was verified for an analytical laminar Couette flow. Subsequently, different laminar and turbulent two-phase flows were used to validate the results against experimental or theoretical data, reaching from the capillary to the gravity scale.
The CH-NS approach naturally includes surface tension effects but the more relevant advantages refer to the efficiency  and the resharpening capability of the approach. Unlike a VoF approach, accurate steady simulations can be performed much faster without any CFL-constraint, since the volume- or mass fraction equation turns into a classical convection-diffusion equation. The latter makes use of taylored, compressive convection schemes obsolet  and supports the use of upwind biased approximation of convective kinematics. 

Efficiency benefits over the traditional VoF method were demonstrated for a fully turbulent two phase flow around a container ship hull at realistic Froude- and large Reynolds-numbers in steady state. For all cases presented in this paper, the computational time is reduced by at least one order of magnitude and minor predictive differences were observed in comparison with the VoF method.

\section{Acknowledgments}
The current work is a part of the research projects "Drag Optimisation of Ship Shapes’" funded by the German Research Foundation (DFG, Grant No. RU 1575/3-1)
as well as "Dynamic Adaptation of Modular Shape Optimization Processes" funded by the German Federal Ministry for Economic Affairs and Energy (BMWi, Grant No.  03SX453B).
This support is gratefully acknowledged by the authors. 
Selected computations were performed with resources provided by the North-German Supercomputing Alliance (HLRN).
In addition, we would like to mention the Matlab Symbolic Toolbox \cite{matlabsymbolic}, which was a great help when calculating the analytical solutions.

\bibliographystyle{plainnat}
\bibliography{./library.bib}
%\bibliography{/home/fds211/kuehl/ownCloud/Publications/Literature/library.bib}

\section{Appendix}

\subsection{Stability Analysis}
\label{app:neumann}
A discrete von-Neumann stability analysis of Equ. \ref{equ:cahn_hilli_diffu}  yields the following amplification factor:
\begin{align}
\xi = \frac{\xi^{m+1}}{\xi^m}= \frac{1 + \mathrm{Co}^\mathrm{diff} \frac{C_\mathrm{2}}{C_1 \delta x^2} \left[4 \mathrm{cos}^2( \varphi ) - 8 \mathrm{cos}(\varphi) + 5 \right]}{1 + \mathrm{Co} \left[ 1 - \mathrm{cos}(\varphi) + i \, \mathrm{sin}(\varphi) \right] + 2 \mathrm{Co}^\mathrm{diff} 
 b^{\prime \prime}\left[ 1 - \mathrm{cos}(\varphi) \right]}
\label{neumann1}
\end{align}
Here $i$ and  $m$  represent the complex constant ($i^2 = -1$) and an  iteration counter, and the parameter $\varphi = \beta \delta x$ with  $\mathrm{cos} \left( \beta \, \delta x \right) = 1/2 \left( e^{-J \, \beta \, \delta x} + e^{J \, \beta \, \delta x} \right)$ represents the phase angle.
The system can be thought of as unconditional stable if $\xi \leq 1$. With attention restricted to the under-resolved situation ($C_\mathrm{1} = \SI{1}{Pa}$, $C_\mathrm{2} = \SI{0}{N}$), we obtain 
\begin{align}
\xi &= \frac{\xi^{m+1}}{\xi^m}= \frac{1}{1 + \left(\mathrm{Co}+ 2 \mathrm{Co}^\mathrm{diff} 
 b^{\prime \prime} \right) \left[ 1 - \mathrm{cos}(\varphi) \right] + i  \; \mathrm{Co} \; \mathrm{sin}(\varphi) }
 \nonumber \\ 
 &= \frac{1 + \left(\mathrm{Co}+ 2 \mathrm{Co}^\mathrm{diff} 
 b^{\prime \prime} \right) \left[ 1 - \mathrm{cos}(\varphi) \right]}{\left(1 + \left(\mathrm{Co}+ 2 \mathrm{Co}^\mathrm{diff} 
 b^{\prime \prime} \right) \left[ 1 - \mathrm{cos}(\varphi) \right]\right)^2 + \left( \mathrm{Co} \; \mathrm{sin}(\varphi)  \right)^2}
- i  \frac{ \mathrm{Co} \; \mathrm{sin}(\varphi)  }{\left(1 + \left(\mathrm{Co}+ 2 \mathrm{Co}^\mathrm{diff} 
 b^{\prime \prime} \right) \left[ 1 - \mathrm{cos}(\varphi) \right]\right)^2 + \left( \mathrm{Co} \; \mathrm{sin}(\varphi)  \right)^2}
\label{neumann2}
\end{align}
Confining the interest to the real term in (\ref{neumann2}),  the system is stable for the pure phases ($c_\mathrm{0} = 0$, $c_\mathrm{0} = 1$).  Along the phase transition regime $c=0.5$ we estimate the stability limits by
\begin{align}
\tilde{M} \leq  \mathrm{1} \ \ \qquad  {\rm and } \qquad  \ \tilde{M} \geq  
\left[1 + \frac{2}{\mathrm{Co} \left[ 1 - \mathrm{cos}(\varphi) \right] } \right]. 
\label{equ:neumann3}
\end{align}

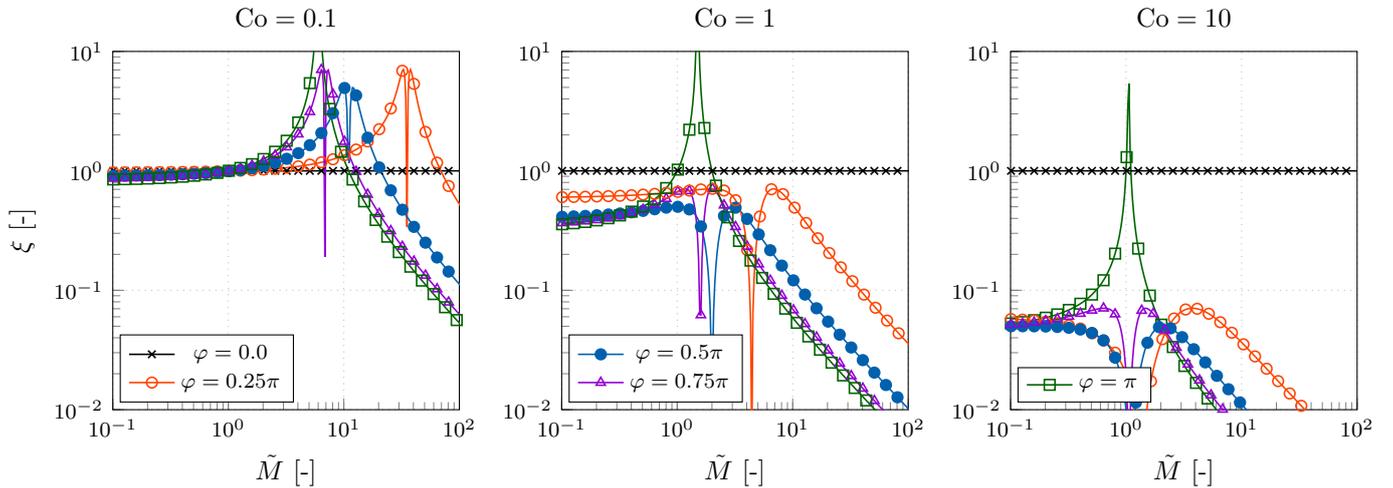
\begin{figure}[h]
\centering
\analytiSolutionPictures
\begin{tikzpicture}
\begin{axis}[
 xlabel style={text width=0.25\textwidth,align=center},
 ylabel style={text width=0.25\textwidth,align=center},
 xlabel={$\tilde{M}$ [-]},
 ylabel={$\xi$ [-]},
 xmode=log,
 ymode=log,
 xmin=1E-01,xmax=1E+02,
 legend style={at={(0.02,0.02)},anchor=south west},
 ymin=1E-02,ymax=1E+01,
 title={$\mathrm{Co} = 0.1$}
]
\addplot [mark1, mark repeat=10] table[x expr={\thisrowno{0}},y expr={\thisrowno{1}}]{data/Stability_Analysis_1E-01.dat};
\addplot [mark2, mark repeat=10] table[x expr={\thisrowno{0}},y expr={\thisrowno{2}}]{data/Stability_Analysis_1E-01.dat};
\addplot [mark3, mark repeat=10] table[x expr={\thisrowno{0}},y expr={\thisrowno{3}}]{data/Stability_Analysis_1E-01.dat};
\addplot [mark4, mark repeat=10] table[x expr={\thisrowno{0}},y expr={\thisrowno{4}}]{data/Stability_Analysis_1E-01.dat};
\addplot [mark5, mark repeat=10] table[x expr={\thisrowno{0}},y expr={\thisrowno{5}}]{data/Stability_Analysis_1E-01.dat};
 
\addlegendentry{$\varphi = 0.0$};
\addlegendentry{$\varphi = 0.25 \pi$};
%\addlegendentry{$\varphi = 0.50 \pi$};
%\addlegendentry{$\varphi = 0.75 \pi$};
%\addlegendentry{$\varphi = 1.00 \pi$};

\end{axis}
\end{tikzpicture}
\begin{tikzpicture}
\begin{axis}[
 xlabel style={text width=0.25\textwidth,align=center},
 ylabel style={text width=0.25\textwidth,align=center},
 xlabel={$\tilde{M}$ [-]},
 xmode=log,
 ymode=log,
 xmin=1E-01,xmax=1E+02,
 legend style={at={(0.02,0.02)},anchor=south west}, 
 ymin=1E-02,ymax=1E+01,
 title={$\mathrm{Co} = 1$}
]
\addplot [mark3, mark repeat=10] table[x expr={\thisrowno{0}},y expr={\thisrowno{3}}]{data/Stability_Analysis_1E-00.dat};
\addplot [mark4, mark repeat=10] table[x expr={\thisrowno{0}},y expr={\thisrowno{4}}]{data/Stability_Analysis_1E-00.dat};
\addplot [mark1, mark repeat=10] table[x expr={\thisrowno{0}},y expr={\thisrowno{1}}]{data/Stability_Analysis_1E-00.dat};
\addplot [mark2, mark repeat=10] table[x expr={\thisrowno{0}},y expr={\thisrowno{2}}]{data/Stability_Analysis_1E-00.dat};
\addplot [mark5, mark repeat=10] table[x expr={\thisrowno{0}},y expr={\thisrowno{5}}]{data/Stability_Analysis_1E-00.dat};

\addlegendentry{$\varphi = 0.5 \pi$};
\addlegendentry{$\varphi = 0.75 \pi$};
%\addlegendentry{$\varphi = 0.00 \pi$};
%\addlegendentry{$\varphi = 0.25 \pi$};
%\addlegendentry{$\varphi = 1.00 \pi$};

\end{axis}
\end{tikzpicture}
\begin{tikzpicture}
\begin{axis}[
 xlabel style={text width=0.25\textwidth,align=center},
 ylabel style={text width=0.25\textwidth,align=center},
 xlabel={$\tilde{M}$ [-]},
 xmode=log,
 ymode=log,
 xmin=1E-01,xmax=1E+02,
 legend style={at={(0.02,0.02)},anchor=south west},
 ymin=1E-02,ymax=1E+01,
 title={$\mathrm{Co} = 10$}
]
\addplot [mark5, mark repeat=10] table[x expr={\thisrowno{0}},y expr={\thisrowno{5}}]{data/Stability_Analysis_1E+01.dat};
\addplot [mark1, mark repeat=10] table[x expr={\thisrowno{0}},y expr={\thisrowno{1}}]{data/Stability_Analysis_1E+01.dat};
\addplot [mark2, mark repeat=10] table[x expr={\thisrowno{0}},y expr={\thisrowno{2}}]{data/Stability_Analysis_1E+01.dat};
\addplot [mark3, mark repeat=10] table[x expr={\thisrowno{0}},y expr={\thisrowno{3}}]{data/Stability_Analysis_1E+01.dat};
\addplot [mark4, mark repeat=10] table[x expr={\thisrowno{0}},y expr={\thisrowno{4}}]{data/Stability_Analysis_1E+01.dat};

\addlegendentry{$\varphi = \pi$};
%\addlegendentry{$\varphi = 0.00 \pi$};
%\addlegendentry{$\varphi = 0.25 \pi$};
%\addlegendentry{$\varphi = 0.50 \pi$};
%\addlegendentry{$\varphi = 0.75 \pi$};

\end{axis}
\end{tikzpicture}
\caption{Amplification factor of a one dimensional CH-NS system over the mobility factor $\tilde{M}$ for different phase angles $\varphi$ and Courant-numbers, i.e.  $\mathrm{Co} = 0.1$ (left), $\mathrm{Co} = 1$ (middle) and $\mathrm{Co} = 10$ (right).}
\label{fig:stability_analysis}
\end{figure}

Fig. \ref{fig:stability_analysis}  depicts the amplification factor for three different Courant-numbers ($\mathrm{Co} = 0.1$, $\mathrm{Co} = 1$, $\mathrm{Co} = 10$) and five different exemplary phase angles as a function of the non-dimensional mobility parameter $\tilde M$. 

\end{document}